\documentclass[a4paper,UKenglish]{lipics-v2018}
 
\usepackage{microtype}
\usepackage{etex}
\usepackage{graphicx} 
\usepackage{microtype}
\usepackage{placeins} 
\usepackage{tikz}
\usepackage{wrapfig}
\usepackage{amsfonts,amsthm,caption,subcaption}
\usetikzlibrary{automata,arrows}
\usepackage{amsmath}
\usepackage{amssymb,latexsym}
\usepackage{mathrsfs}
\usepackage{stmaryrd}
\usepackage{verbatim}
\usepackage{thm-restate}
\usepackage[utf8]{inputenc} 
\usepackage[T1]{fontenc}
\usepackage[colorinlistoftodos,
textsize=footnotesize,color=orange!70]{todonotes}
\usepackage[arrow,matrix,frame,curve,cmtip]{xy}\CompileMatrices

\newcommand{\short}[1]{} 
\newcommand{\full}[1]{#1}

\newcommand{\Set}{\mathbf{Set}}
\newcommand{\PMet}{\mathbf{PMet}}
\newcommand{\DPMet}{\mathbf{DPMet}}

\renewcommand{\epsilon}{\varepsilon}
\renewcommand{\phi}{\varphi}
\newcommand{\myrightarrow}[1]{\mathrel{\raisebox{-1pt}{$\xrightarrow{#1}$}}}
\newcommand{\neto}{\myrightarrow{1}}

\theoremstyle{plain}
\newtheorem{prop}{Proposition}
\newtheorem{lem}[prop]{Lemma}

\newtheorem{defi}[prop]{Definition}
\newtheorem{ex}[prop]{Example}
\nolinenumbers

\full{\hideLIPIcs}

\title{(Metric) Bisimulation Games and Real-Valued Modal Logics for
	Coalgebras}

\titlerunning{(Metric) Bisimulation Games and Real-Valued Modal Logics for
	Coalgebras}

\author{Barbara K\"onig }{Universität Duisburg-Essen,
	Germany}{barbara\_koenig@uni-due.de}{}{}{}
\author{Christina Mika-Michalski }{Universität Duisburg-Essen,
	Germany}{christina.mika-michalski@uni-due.de}{}{}{}

\authorrunning{B. K\"onig and C. Mika-Michalski} 

\Copyright{B. K\"onig and C. Mika-Michalski}

\subjclass{, }


\category{}

\relatedversion{}

\supplement{}

\funding{Research partially supported by DFG Project BEMEGA.}

\acknowledgements{ We thank Paul Wild, Lutz Schr\"oder and Dirk
  Pattinson for inspiring discussions on fuzzy modal logic, and in
  particular on preservation of total boundedness.}

\EventEditors{Sven Schewe and Lijun Zhang} 
\EventNoEds{2} 
\EventLongTitle{29th International Conference on Concurrency Theory (CONCUR 2018)} \EventShortTitle{CONCUR 2018} 
\EventAcronym{CONCUR} 
\EventYear{2018} 
\EventDate{September 4--7, 2018} 
\EventLocation{Beijing, China}
\EventLogo{} 
\SeriesVolume{118} \ArticleNo{37} 

\begin{document}

\maketitle

\begin{abstract}
  Behavioural equivalences can be characterized via bisimulations,
  modal logics and spoiler-defender games. In this paper we review
  these three perspectives in a coalgebraic setting, which allows us
  to generalize from the particular branching type of a transition
  system. We are interested in qualitative notions (classical
  bisimulation) as well as quantitative notions (bisimulation
  metrics).
  
  Our first contribution is to introduce a spoiler-defender
  bisimulation game for coalgebras in the classical case. Second, we
  introduce such games for the metric case and furthermore define a
  real-valued modal coalgebraic logic, from which we can derive the
  strategy of the spoiler. For this logic we show a quantitative
  version of the Hennessy-Milner theorem.
\end{abstract}

\section{Introduction}
\label{sec:introduction}
In the characterization of behavioural equivalences one encounters the
following triad: First, such equivalences can be described via
bisimulation relations, where the largest bisimulation (or
bisimilarity) can be characterized as a greatest fixpoint. Second, a
modal logic provides us with bisimulation-invariant formulas and the
aim is to prove a Hennessy-Milner theorem which says that two states
are behaviourally equivalent if and only if they satisfy the same
formulas \cite{hm:observing-nondet-conc}. A third, complementary view
is given by spoiler-defender games \cite{Stirling1999}. Such games are
useful both for theoretical reasons, see for instance the role of
games in the Van Benthem/Rosen theorem
\cite{o:van-Benthem-Rosen-elementary}, or for didactical purposes, in
particular for showing that two states are not behaviourally
equivalent. The game starts with two tokens on two states and the
spoiler tries to make a move that cannot be imitated by the
defender. If the defender is always able to match the move of the
spoiler we can infer that the two initial states are behaviourally
equivalent.  If the states are not equivalent, a strategy for the
spoiler can be derived from a distinguishing modal logic formula.

Such games are common for standard labelled transition systems, but
have been studied for other types of transition systems only to a
lesser extent. For probabilistic transition systems there are
  game characterizations in
  \cite{Desharnais,fkp:expressiveness-prob-modal-logics}, where the
  players can make moves to sets of states, rather than take a
  transition to a single state. Furthermore, in
\cite{cd:game-process-equivalences} a general theory of games is
introduced in order to characterize process equivalences of the
linear/branching time spectrum.

Our aim is to extend this triad of bisimulation, logics and games in
two orthogonal dimensions. First, we work in the general framework of
coalgebras \cite{r:universal-coalgebra}, which allows to specify and
uniformly reason about systems of different branching types
(e.g. non-deterministic, probabilistic or weighted), parameterized
over a functor. While behavioural equivalences
\cite{s:relating-coalgebraic-bisimulation} and modal logics
\cite{SCHRODER2008230,PATTINSON2003177} have been extensively studied
in this setting, there are almost no contributions when it comes to
games. We are mainly aware of the work by Baltag \cite{clg:Baltag},
which describes a coalgebraic game based on the bisimulation relation,
which differs from the games studied in this paper and is associated
with another variant of logic, namely Moss' coalgebraic logics
\cite{m:coalgebraic-logic}. A variant of Baltag's game was used in
\cite{k:terminal-sequence-games} for terminal sequence induction via
games. (There are more contributions on evaluation games which
describe the evaluation of a modal formula on a transition system, see
for instance \cite{flv:coalgebra-automata}.) Our contribution
generalizes the games of \cite{Desharnais} and allows us, given a new
type of system characterized by a functor on the category $\Set$,
satisfying some mild conditions, to automatically derive the
corresponding game. The second dimension in which we generalize is to
move from a qualitative to a quantitative notion of behavioural
equivalence. That is, we refrain from classifying systems as either
equivalent or non-equivalent, which is often too strict, but rather
measure their behavioural distance. This makes sense in probabilistic
systems, systems with time or real-valued output. For instance, we
might obtain the result, that the running times of two systems differ
by 10~seconds, which might be acceptable in some scenarios (departure
of a train), but inacceptable in others (delay of a vending
machine). On the other hand, two states are behaviourally equivalent
in the classical sense if and only if they have distance~$0$. Such
notions are for instance useful in the area of conformance testing
\cite{Khakpour2015NotionsOC} and differential privacy
\cite{cgpx:generalized-bisim-metrics}.

Behavioural metrics have been studied in different variants, for
instance in probabilistic settings
\cite{d:labelled-markov-processes,dgjp:metrics-labelled-markov,cgt:logical-bisim-metrics}
as well as in the setting of metric transition systems
\cite{deAlfaro:2009:LBS:1525651.1525793,flt:quantitative-spectrum},
which are non-deterministic transition systems with quantitative
information. The groundwork for the treatment of coalgebras in metric
spaces was laid by Turi and Rutten \cite{tr:final-metric}. We showed
how to characterize behavioural metrics in coalgebras by studying
various possibilities to lift functors from $\Set$ to the category of
(pseudo-)metric spaces
\cite{BBKK15b,bbkk:coalgebraic-behavioral-metrics}. Different from
\cite{tr:final-metric,VANBREUGEL2005115} we do not assume that the
coalgebra is given a priori in the category of pseudometric spaces,
that is we have to first choose a lifting of the behaviour functor in
order to specify the behavioural metric. Such liftings are not
unique\footnote{In fact, consider the product bifunctor
  $F(X,Y) = X\times Y$, for which there are several liftings: we can
  e.g. use the maximum or the sum metric. While the maximum metric is
  canonically induced by the categorical product, the sum metric is
  also fairly natural.} and in particular we introduced in
\cite{BBKK15b,bbkk:coalgebraic-behavioral-metrics} the Kantorovich and
the Wasserstein liftings, which generalize well-known liftings for the
probabilistic case and also capture the Hausdorff metric. Here we use
the Kantorovich lifting, since this lifting integrates better with
coalgebraic logic. Our results are parameterized over the lifting, in
particular the behavioural metrics, the game and the logics are
dependent on a set $\Gamma$ of evaluation functions.

In the metric setting it is natural to generalize from classical
two-valued logics to real-valued modal logics and to state a
corresponding Hennessy-Milner theorem that compares the behavioural
distance of two states with the logical distance, i.e., the supremum
of the differences of values, obtained by the evaluation of all
formulas.  Such a Hennessy-Milner theorem for probabilistic transition
systems was shown in \cite{dgjp:metrics-labelled-markov} and also
studied in a coalgebraic setting
\cite{VANBREUGEL2005115,bw:behavioural-distances}. Similar results
were obtained in \cite{wspk:van-benthem-fuzzy} for fuzzy logics, on
the way to proving a van Benthem theorem. Fuzzy logics were also
studied in \cite{sp:descr-logics-fuzzy} in a general coalgebraic
setting, but without stating a Hennessy-Milner theorem.

Here we present a real-valued coalgebraic modal logic and give a
Hennessy-Milner theorem for the general coalgebraic setting as a new
contribution. Our proof strategy follows the one for the probabilistic
case in \cite{VANBREUGEL2005115}. We need several concepts from real
analysis, such as non-expansiveness and total boundedness in order to
show that the behavioural distance (characterized via a fixpoint) and
the logical distance coincide.

Furthermore we give a game characterization of this behavioural metric
in a game where we aim to show that $d(x,y)\le \epsilon$, i.e., the
behavioural distance of two states $x,y$ is bounded by $\epsilon$.
Furthermore, we work out the strategies for the defender and spoiler:
While the strategy of the defender is based on the knowledge of the
behavioural metric, the strategy of the spoiler can be derived from a
logical formula that distinguishes both states.

Again, work on games is scarce: \cite{Desharnais} presents a game
which characterizes behavioural distances, but pairs it with a
classical logic.

The paper is organized as follows: we will first treat the classical
case in Section~\ref{sec:classical}, followed by the metric case in
Section~\ref{sec:metric}. The development in the metric case is more
complex, but in several respects mimics the classical case. Hence, in
order to emphasize the similarities, we will use the same structure
within both sections: we start with foundations, followed by the
introduction of modal logics and the proof of the Hennessy-Milner
theorem. Then we will introduce the game with a proof of its soundness
and completeness. Finally we will show how the strategy for the
spoiler can be derived from a logical formula. In the end we wrap
everything up in the conclusion
(Section~\ref{sec:conclusion}). \full{The proofs can be found in
  Appendix~\ref{sec:proofs}.}\short{The proofs can be found in the
  full version of this paper
  \cite{km:bisim-games-logics-metric-arxiv}.}

\section{Logics and Games for the Classical Case}
\label{sec:classical}

\subsection{Foundations for the Classical Case}
\label{sec:foundations-classical}

We fix an endofunctor $F\colon \Set\to \Set$, intuitively describing
the branching type of the transition system under consideration. A
\emph{coalgebra}, describing a transition system of this branching
type is given by a function $\alpha\colon X\to FX$
\cite{r:universal-coalgebra}.  Two states $x,y\in X$ are considered to
be \emph{behaviourally equivalent} ($x\sim y$) if there exists a
coalgebra homomorphism $f$ from $\alpha$ to some coalgebra
$\beta\colon Y\to FY$ (i.e., a function $f\colon X\to Y$ with
$\beta\circ f = Ff\circ \alpha$) such that $f(x) = f(y)$.

\begin{ex}
  \label{ex:prob-ts}
  We consider the (finitely or countably supported) probability
  distribution functor $\mathcal{D}$ with
  $\mathcal{D}X = \{p\colon X\to [0,1]\mid \sum_{x\in X} p(x) = 1\}$
  (where the $p$ are either finitely or countable
  supported). Furthermore let $1 = \{\bullet\}$ be a singleton set.
  
  Now coalgebras of the form $\alpha\colon X\to \mathcal{D}X+1$ can be
  seen as probabilistic transition systems where each state $x$ is
  either terminating ($\alpha(x) = \bullet$) or is associated with a
  probability distribution on its successor states. Note that one
  could easily integrate labels as well, but we will work with this
  version for simplicity.
  
  Figure~\ref{fig:prob-ts} displays an example coalgebra (where
  $0\le\epsilon\le \frac{1}{2}$). Note that whenever $\epsilon = 0$,
  we have $x\sim y$, since there is a coalgebra homomorphism from the
  entire state space into the right-hand side component of the
  transition system. If $\epsilon > 0$ we have $1\not\sim 2$.
\end{ex}

\begin{figure}[h]
  \begin{subfigure}[t]{0.55\textwidth}
    \centering
    \begin{tikzpicture}
      \node (S) at (0,-1.5) {$x$}; 			
      \node (S1) at (0,-2) [circle,draw]{$1$}; 			
      \node (S3) at (-1.6,-3.2) [circle,draw]{$3$};
      \node [accepting](S4) at (0,-3.5) [circle,draw]{$4$};  
      \node [accepting](S5) at (1.6,-3.2) [circle,draw]{$5$}; 	
      \draw  [->] (S1) to node [above]{$\frac{1}{2}$} (S3);
      \draw  [->] (S1) to node [above]{$\frac{1}{4}$} (S5);	
      \draw  [->] (S1) to node [right]{$\frac{1}{4}$} (S4);		
      \path
      (S3) edge [loop below] node {$1$} (S3);			 
      \node (T) at (4,-1.5) {$y$}; 			
      \node (T1) at (4,-2) [circle,draw]{$2$}; 			
      \node (T6) at (2.8,-3.2) [circle,draw]{$6$};  
      \node [accepting](T7) at (5.2,-3.2) [circle,draw]{$7$}; 
      \node (t1) at (2.9,-2.4) {$\frac{1}{2}-\epsilon$};
      \node (t2) at (5.2,-2.4) { $\frac{1}{2}+\epsilon$};
      
      \draw  [->] (T1) to node {} (T6);
      \draw  [->] (T1) to node {} (T7);		
      \path
      (T6) edge [loop below] node {$1$} (T6);	
    \end{tikzpicture}		
    \subcaption{Probabilistic transition system for the functor 
      $FX = \mathcal{D}X+1$}	
    \label{fig:prob-ts}			
  \end{subfigure}%
  \hfill
  \begin{subfigure}[t]{0.4\textwidth}
    \centering
    \begin{tikzpicture}
      \node (S) at (0,-1.5) {$x$}; 			
      \node (S1) at (0,-2) [circle,draw]{$1$}; 			
      \node (S3) at (-0.8,-3.2) [circle,draw]{$3$};
      \node (S4) at (0.8,-3.2) [circle,draw]{$4$}; 
      \draw  [->] (S1) to node [left]{$a$} (S3);
      \draw  [->] (S1) to node [right]{$a$} (S4);	
      \node (S) at (3,-1.5) {$y$}; 			
      \node (T2) at (3,-2) [circle,draw]{$2$}; 			
      \node (T5) at (3,-3.2) [circle,draw]{$5$};  
      \node (S6) at (-0.8,-4.4) [circle,draw]{$6$}; 			
      \node (S7) at (0.8,-4.4) [circle,draw]{$7$};  
      \node (T8) at (2.2,-4.4) [circle,draw]{$8$}; 			
      \node (T9) at (3.8,-4.4) [circle,draw]{$9$};  
      \draw  [->] (T2) to node [left]{$a$} (T5);
      \draw  [->] (S3) to node [left]{$b$} (S6);
      \draw  [->] (S4) to node [right]{$c$} (S7);
      \draw  [->] (T5) to node [left]{$b$} (T8);
      \draw  [->] (T5) to node [right]{$c$} (T9);
    \end{tikzpicture}		
    \subcaption{Non-deterministic transition system for the functor
      $FX = \mathcal{P}(A\times X)$}
    \label{fig:nondet-ts}
  \end{subfigure}
  \caption{}
\end{figure}

Furthermore we need the lifting of a preorder under a functor $F$. For
this we use the lifting introduced in
\cite{bk:finitary-functors-set-preord-poset} which guarantees that
the lifted relation is again a preorder whenever $F$ preserves weak
pullbacks: Let $ \leq $ be a preorder on $Y$, i.e.
$ \leq \ \subseteq Y \times Y $. We define the preorder
$ \leq^F \subseteq FY \times FY$ with $ t_1, t_2 \in FY $ as follows:
$ t_1 \leq^F t_2 $ whenever some $ t \in \ F(\leq) $ exists such that
$ F\pi_i(t) = t_i$, where $ \pi_i \colon \leq\ \to Y$ with
$ i \in \{1,2\} $ are the usual projections.

\begin{restatable}{lem}{LemLiftOrderPred}
  \label{lem:lift-order-pred}
  Let $(Y,\le)$ be an ordered set and let $p_1,p_2\colon X\to Y$ be
  two functions. Then $ p_{1} \leq p_{2} $ implies
  $Fp_{1} \leq^{F} Fp_{2} $. (Both inequalities are to be read as
  pointwise ordering.)
\end{restatable}

\begin{ex}
  \label{ex:lifted-order}
  We are in particular interested in lifting the order $0\le 1$ on
  $2 = \{0,1\}$. In the case of the distribution functor $\mathcal{D}$
  we have $\mathcal{D}2 \cong [0,1]$ and $\le^\mathcal{D}$ corresponds
  to the order on the reals. For the powerset functor $\mathcal{P}$ we
  obtain the order $\{0\}\le^\mathcal{P}\{0,1\}\le^\mathcal{P} \{1\}$
  where $\emptyset$ is only related to itself.
\end{ex}

\subsection{Modal Logics  for the Classical Case}
\label{sec:logics-classical}

We will first review coalgebraic modal logics, following mainly
\cite{PATTINSON2003177,SCHRODER2008230}, using slightly different, but
equivalent notions. In particular we will introduce a logic where a
predicate lifting is given by an evaluation map of the form
$\lambda\colon F2\to 2$, rather than by a natural transformation. In
particular, each predicate $p\colon X\to 2$ is lifted to a predicate
$\lambda\circ Fp\colon FX\to 2$. We do this to obtain a uniform
presentation of the material.  Of course, both views are equivalent,
as spelled out in \cite{SCHRODER2008230}.

Given a cardinal $\kappa$ and a set $\Lambda$ of evaluation maps
$\lambda\colon F2\to 2$, we define a coalgebraic modal logic
$\mathcal{L}^{\kappa}(\Lambda)$ via the grammar:
\[ \phi ::= \bigwedge \Phi \mid \neg \phi \mid [\lambda] \phi
  \quad\text{where $\Phi \subseteq \mathcal{L}^{\kappa}(\Lambda)$ with
    $\mathit{card}(\Phi) < \kappa$ and $\lambda \in \Lambda$.} \]
The last case describes the prefixing of a formula $\phi$ with a
modality $[\lambda]$.

Given a coalgebra $\alpha\colon X\to FX$ and a formula $\phi$, the
semantics of such a formula is given by a map
$\llbracket\phi\rrbracket_\alpha\colon X\to 2$, where conjunction and
negation are interpreted as usual and
$\llbracket [\lambda]\phi \rrbracket_{\alpha}=\lambda\circ
F\llbracket\phi\rrbracket_\alpha\circ \alpha$. For simplicity we will
often write $\llbracket \phi\rrbracket$ instead of
$\llbracket \phi\rrbracket_\alpha$. Furthermore for $x\in X$ we write
$x\models \phi$ whenever
$\llbracket \phi \rrbracket_\alpha(x) = 1$.

In \cite{PATTINSON2003177} Pattinson has isolated the property of a
separating set of predicate liftings to ensure that logical and
behavioural equivalence coincide, i.e., the Hennessy-Milner property
holds. It means that every $t \in FX$ is uniquely determined by the
set
$ \{(\lambda, p) \mid \lambda\in\Lambda, p\colon X\to 2,
\lambda(Fp(t)) = 1 \} $.

\begin{defi}
  \label{def:separating-functor}
  A set $\Lambda$ of evaluation maps is separating for a functor
  $F\colon \Set\to \Set$ whenever for all sets $X$ and $t_1,t_2\in FX$ with $t_1 \neq t_2$
  there exists $\lambda\in \Lambda$ and $p\colon X\to 2$ such that
  $\lambda(Fp(t_1)) \neq \lambda(Fp(t_2))$.
\end{defi}

The Hennessy-Milner theorem for coalgebraic modal
  logics can be stated as follows. The result has
  already been presented in
  \cite{PATTINSON2003177,SCHRODER2008230,g:universal-coalgebras-logics}\full{,
    but for completeness we give a proof based on
    evaluation maps in the appendix.}  

\begin{restatable}[\cite{SCHRODER2008230}]{prop}{PropHmLogic}
  \label{prop:hm-logic}
  The logic $\mathcal{L}^{\kappa}(\Lambda)$ is sound, that is given a
  coalgebra $\alpha\colon X\to FX$ and $x,y\in X$, $x\sim y$ implies
  that
  $\llbracket \phi \rrbracket_\alpha(x) = \llbracket \phi
  \rrbracket_\alpha(y)$ for all formulas $\phi$.
  
  Whenever $F$ is $\kappa$-accessible\footnote{A functor
    $F\colon \Set \to \Set$ is $\kappa$-accessible if for all sets $X$
    and all $x\in FX$ there exists $Z\subseteq X$, $|Z| < \kappa$ such
    that $x \in FZ \subseteq FX$
    \cite{agt:presentation-set-functors}. (Note that we use the
      fact that $\Set$-functors preserve injections $f\colon A\to B$
      whenever $A\neq \emptyset$.) For details and a more categorical
    treatment see \cite{ar:locally-presentable-categories}.}  and
  $\Lambda$ is separating for $F$, the logic is also expressive:
  whenever
  $\llbracket \phi \rrbracket_\alpha(x) = \llbracket \phi
  \rrbracket_\alpha(y)$ for all formulas $\phi$ we have that
  $x\sim y$.
\end{restatable}

In \cite{SCHRODER2008230} it has been shown that a functor $F$ has a
separating set of predicate liftings iff
$(Fp \colon FX \rightarrow F2)_{p\colon X \rightarrow 2}$ is jointly
injective. We extend this characterization to monotone predicate
liftings, respectively evaluation maps, i.e., order-preserving maps
$\lambda\colon (F2,\le^F) \to (2,\le)$ where $\le$ is the order
$0\le 1$. This result will play a role in 
Section~\ref{sec:games-classical}.

\begin{restatable}{prop}{PropSepPlAsJi}
  \label{prop:separating-pl-antisym-ji}
  $F$ has a separating set of monotone evaluation maps iff $\leq^F$ is
  anti-symmetric and
  $(Fp \colon FX \rightarrow F2)_{p\colon X \rightarrow 2}$ is jointly
  injective.
\end{restatable}

Note that an evaluation map is monotone if and only if its induced
predicate lifting is monotone \full{(see
  Proposition~\ref{prop:ev-monotone} in the appendix).}\short{(see
  \cite{km:bisim-games-logics-metric-arxiv}).}

\subsection{Games for the Classical Case}
\label{sec:games-classical}

We will now present the rules for the behavioural equivalence game. At
the beginning of a game, there are two states $x,y$ available for
selection. The aim of the spoiler (S) is to prove that $x\not\sim y$,
the defender (D) attempts to show the opposite.

\begin{itemize}
  \itemsep0pt
\item \textbf{Initial situation:} Given a coalgebra
  $ \alpha \colon X \to FX $, we start with $ x,y \in X $.
\item \textbf{Step~1:} S chooses $s\in \{x,y\}$ and a predicate
  $ p_{1} \colon X \to 2 $.
\item \textbf{Step~2:} D takes $t\in\{x,y\}\backslash\{s\}$ and
  has to answer with a predicate
  $ p_{2} \colon X \to 2 $, which satisfies
  $ Fp_{1}(\alpha(s)) \leq^{F} Fp_{2}(\alpha(t)) $.
\item \textbf{Step~3:} S chooses $p_{i} $ with $i \in \{1,2\}$
  and some state $ x' \in X$ with $ p_{i}(x')=1$.
\item \textbf{Step~4:} D chooses some state $ y' \in X$ with
  $ p_{j}(y')=1$ where $i\neq j$.
\end{itemize}

After one round the game continues in Step~1 with states $ x'$ and
$y'$. D wins the game if the game continues forever or if S has no
move at Step~3. In the other cases, i.e. D has no move at Step~2 or
Step~4, S wins.

In a sense such a game seems to mimic the evaluation of a modal
formula, but note that the chosen predicates do not have to be
bisimulation-invariant, as opposed to modal formulas.

\begin{restatable}{theo}{ThmWinningStrategyDefender}
  \label{thm:winning-strategy-defender}
  Assume that $F$ preserves weak pullbacks and has a separating set of
  monotone evaluation maps. Then $ x \sim y $ iff D has
  winning strategy for the initial situation $ (x,y) $.
\end{restatable}

Part of the proof of Theorem~\ref{thm:winning-strategy-defender} is to
establish a winning strategy for D whenever $x\sim y$. We will quickly
sketch this strategy: In Step~1 S plays $p_1$ which represents a set
of states. One good strategy for D in Step~2 is to close this set
under behavioural equivalence, i.e., to add all states which are
behaviourally equivalent to a state in $p_1$, thus obtaining $p_2$. It
can be shown that $ Fp_{1}(\alpha(s)) \leq^{F} Fp_{2}(\alpha(t)) $
always holds for this choice. Now, if S chooses $x',p_1$ in Step~3, D
simply takes $x'$ as well. On the other hand, if S chooses $x',p_2$,
then either $x'$ is already present in $p_1$ or a state $y'$ with
$x'\sim y'$. D simply chooses $y'$ and the game continues.

\begin{ex}
  \label{ex:nondet-ts}
  Now consider an example coalgebra for the functor
  $FX = \mathcal{P}(A\times X)$, where $\mathcal{P}$ is the powerset
  functor (see Figure~\ref{fig:nondet-ts}). Obviously $x\not\sim y$,
  so S must have a winning strategy. Somewhat different from the usual
  bisimulation game, here the two players play subsets of the state
  space, instead of single states. Otherwise the game proceeds
  similarly.
  
  Assume that S chooses $s = 1$ and defines a predicate $p_1$, which
  corresponds to the set $\{3\}$. Then $Fp_1(\alpha(s))$ is
  $\{(a,0),(a,1)\}$ (one $a$-successor of $s$ -- $3$ -- satisfies the
  predicate, the other -- $4$ -- does not). Now D must take $t = 2$
  and has to choose a predicate $p_2$ where at least $p_2(5) = 1$. In
  this case $Fp_2(\alpha(t))$ is $\{(a,1)\}$ and $\{1\}$ is larger
  than $\{0,1\}$ in the lifted order (see
  Example~\ref{ex:lifted-order}). However, now S can pick $5$, which
  leaves only $3$ to D.
  
  In the next step, S can choose $s=5$ and a predicate $p_1$, which
  corresponds to $\{9\}$. Hence $Fp_1(\alpha(s))$ is
  $\{(b,0),(c,1)\}$, but it is impossible for D to match this, since
  $(c,1)$ is never contained in $Fp_2(\alpha(t))$ for $t = 3$.
  
  We can see from this game why it is necessary to use an inequality
  $\le^F$ instead of an equality. If there were no $b,c$-transitions
  (just $a$-transitions), $1\sim 2$ would hold.  And then, as
  explained above, D cannot match the move of S exactly, but only by
  choosing a larger value.
\end{ex}
\full{This game is inspired by the game for labelled Markov processes
  in \cite{Desharnais} and the connection is explained in more detail
  in Appendix~\ref{sec:comparison-desharnais}.}\short{This game is
  inspired by the game for labelled Markov processes in
  \cite{Desharnais} and the connection is explained in more detail in
  \cite{km:bisim-games-logics-metric-arxiv}.}

Note that in the probabilistic version of the game, it can again be
easily seen that an inequality is necessary in Step~2: if, in the
system in Figure~\ref{fig:prob-ts} (where $\epsilon = 0$), S chooses
$s = 1$ and $p_1$ corresponds to $\{4\}$, then D can only answer by
going to $7$, which results in a strictly larger value. That is, we
must allow D to do ``more'' than S.

\paragraph*{Game variant:} By looking at the proof of
Theorem~\ref{thm:winning-strategy-defender} it can be easily seen that
the game works as well if we replace the condition
$Fp_1(\alpha(s))\le^F Fp_2(\alpha(t))$ in Step~2 by
$\lambda(Fp_1(\alpha(s)))\le \lambda(Fp_2(\alpha(t)))$ for all
$\lambda\in \Lambda$, provided that $\Lambda$ is a separating set of
monotone evaluation maps. This variant is in some ways less desirable,
since we have to find such a set $\Lambda$ (instead of simply
requiring that it exists), on the other hand in this case the proof
does not require weak pullback preservation, since we do not any more
require transitivity of $\le^F$. This variant of the game is
conceptually quite close to the $\Lambda$-bisimulations studied in
\cite{gs:sim-bisim-coalg-logic}.  In our notation, a relation
$S\subseteq X \times X$ is a $\Lambda$-bisimulation, if whenever
$x\,S\,y$, then for all $\lambda\in\Lambda$, $p\colon X\to 2$,
$\lambda(Fp(\alpha(x)))\le \lambda(Fq(\alpha(y)))$, where $q$
corresponds to the image of $p$ under $S$ (and the same condition
holds for $S^{-1}$). $\Lambda$-bisimulation is sound and complete for
behavioural equivalence if $F$ admits a separating set of monotone
predicate liftings, which coincides with our condition.

\subsection{Spoiler Strategy for the Classical Case}
\label{sec:spoiler-classical}

In bisimulation games the winning strategy for~D can be derived from
the bisimulation or, in our case, from the map $f$ that witnesses the
behavioural equivalence of two states $x,y$ (see the remark after
Theorem~\ref{thm:winning-strategy-defender}). Here we will show that
the winning strategy for~S can be derived from a modal formula $\phi$
which distinguishes $x,y$, i.e., $x\models \phi$ and
$y\not\models \phi$. We assume that all modalities are monotone
(cf. Proposition~\ref{prop:separating-pl-antisym-ji}).

The spoiler strategy is defined over the structure of $\phi$:

\begin{itemize}
\item $\phi = \bigwedge \Phi$: in this case S picks a formula
  $\psi\in\Phi$ with $y\not\models\psi$.
\item $\phi = \lnot \psi$: in this case $S$ takes $\psi$ and reverses
  the roles of $x,y$.
\item $\phi = [\lambda] \psi$: in this case S chooses $x$ and
  $p_1 = \llbracket \psi\rrbracket$ in Step~1. After D has chosen $y$
  and some predicate $p_2$ in Step~2, we can observe that
  $p_2\not\le \llbracket\psi\rrbracket$ \short{(see proof of
  Theorem~\ref{thm:winning-strategy-spoiler} in \cite{km:bisim-games-logics-metric-arxiv})}\full{(see proof of
  Theorem~\ref{thm:winning-strategy-spoiler})}. Now in Step~3 S chooses
  $p_2$ and a state $y'$ with $p_2(y')=1$ and $y'\not\models
  \psi$. Then D must choose $\llbracket\psi\rrbracket$ and a state
  $x'$ with $x'\models \psi$ in Step~4 and the game continues with
  $x',y'$ and $\psi$.
\end{itemize}

It can be shown that this strategy is successful for the spoiler.

\begin{restatable}{theo}{ThmWinningStrategySpoiler}
  \label{thm:winning-strategy-spoiler}
  Assume that $\alpha\colon X\to FX$ is a coalgebra and $F$ satisfies
  the requirements of Theorem~\ref{thm:winning-strategy-defender}.
  Let $\phi$ be a formula that contains only monotone modalities and
  let $x\models \phi$ and $y\not\models \phi$. Then the spoiler
  strategy described above is winning for $S$.
\end{restatable}

\section{Logics and Games for the Metric Case}
\label{sec:metric}

We will now consider the metric version of behavioural equivalence,
including logics and games. Analogous to Section~\ref{sec:classical}
we will first introduce behavioural distance, which will in this case
be defined via the Kantorovich lifting and is parameterized over a set
$\Gamma$ of evaluation maps. Then we introduce a modal logic inspired
by \cite{bw:behavioural-distances} and show a quantitative coalgebraic
analogue of the Hennessy-Milner theorem. This leads us to the
definition of a game for the metric case, where we prove that the
distance induced by the game coincides with the behavioural
distance. We will conclude by explaining how the strategy for the
spoiler can be derived from a logical formula distinguishing two states.

\subsection{Foundations for the Metric Case}
\label{sec:foundations-metric}

Note that this subsection contains several results which are new with
respect to \cite{bbkk:coalgebraic-behavioral-metrics}, in particular
the extension of the Kantorovich lifting to several evaluation maps
and Propositions~\ref{prop:local-nonexpansive},
\ref{prop:lifting-nonexpansive},
\ref{prop:ev-nonexpansive-sup-continuous},
\ref{prop:comp-evaluation-maps} and~\ref{prop:fixpoint-omega-steps}.

In the following we assume that $\top$ is an element of
$\mathbb{R}_0$, it denotes the upper bound of our distances.

We first define the standard notions of pseudometric space and
non-expansive functions.

\begin{defi}[Pseudometric, pseudometric space]
  Let $X$ be a set and $d\colon X \times X \rightarrow [0, \top]$ a
  real-valued function, we call $d$ a pseudometric if it satisfies
  \begin{enumerate}
  \item \label{cond:reflexive} $d(x,x)=0$ ($d$ is a metric if in
    addition $d(x,y) = 0$ implies $x=y$.)
  \item \label{cond:symmetric} $d(x,y) = d(y,x)$
  \item \label{cond:triangle} $d(x,z) \leq d(x,y) + d(y,z)$
  \end{enumerate}
  for all $x,y,z \in X$. If $d$ satisfies only
  Condition~\ref{cond:reflexive} and~\ref{cond:triangle}, it is a
  directed pseudometric.
  
  A (directed) pseudometric space is a pair $(X, d)$ where $X$ is a
  set and $d$ is a (directed) pseudometric on $X$.
\end{defi}

\begin{ex}
  We will consider the following (directed) metrics on $[0,\top]$: the
  Euclidean distance
  $d_e:[0,\top] \times [0,\top] \rightarrow [0,\top]$ with
  $ d_e(a,b)= |a-b|$ and truncated subtraction with
  $d_\ominus(a,b) = a\ominus b = \max\{a-b,0\}$. Note that
  $d_e(a,b) = \max\{d_\ominus(a,b),d_\ominus(b,a)\}$.
\end{ex}

Maps between pseudometric spaces are given by non-expansive functions,
which guarantee that mapping two elements either preserves or
decreases their distance.

\begin{defi}[Non-expansive function]
  Let $(X, d_X)$, $(Y, d_Y)$ be pseudometric spaces. A function
  $f \colon X \rightarrow Y$ is called non-expansive if
  $d_X(x,y) \geq d_Y(f(x),f(y))$ for all $x,y \in X$. In this
  case we write $f\colon (X,d_X)\neto (Y,d_Y)$.
  By $\PMet$ ($\DPMet$) we denote the category of (directed)
  pseudometric spaces and non-expansive functions.
\end{defi}

On some occasions we need to transform an arbitrary function into a
non-expansive function, which can be done as follows.

\begin{restatable}{lem}{LemApproxNonexpansive}
  \label{lem:approx-nonexpansive}
  Let $d$ be a pseudometric on $X$ and let
  $f\colon X\rightarrow [0,\top]$ be any function. Then the function
  $h\colon (X,d)\to ([0,\top],d_e)$ defined via
  $h(z) = \sup\{f(u)-d(u,z) \mid u \in X\}$ is non-expansive and
  satisfies $f \le h$.
  
  Analogously the function $g\colon (X,d)\to ([0,\top],d_e)$ defined
  via $g(z) = \inf\{f(u)+d(u,z) \mid u \in X\}$ is non-expansive and
  satisfies $g\le f$.
\end{restatable}

We will now define the Kantorovich lifting for $\Set$-functors,
introduced in \cite{BBKK15b}. Given a functor $F$ we lift it to a
functor $\bar{F}\colon \PMet\to \PMet$ such that $UF = \bar{F}U$,
where $U$ is the forgetful functor, discarding the pseudometric. The
Kantorovich lifting is parameterized over a finite set $\Gamma$ of
evaluation maps $\gamma \colon F[0,\top]\to [0,\top]$, the analogue to
the evaluation maps for modalities in the classical case. This is an
extension of the lifting in \cite{BBKK15b} where we considered only a
single evaluation map. The new version allows to capture additional
examples, without going via the somewhat cumbersome multifunctor
lifting described in \cite{BBKK15b}.

\begin{defi}[Kantorovich lifting]
  \label{def:kantorovich-lifting}
  Let $F$ be an endofunctor on $\Set$ and let $\Gamma$ be a finite set
  of evaluation maps $\gamma\colon F[0,\top] \rightarrow [0,\top]$.
  For every pseudometric space $(X,d)$ the Kantorovich pseudometric on
  $FX$ is the function
  $d^{\uparrow \Gamma}\colon FX \times FX \rightarrow [0,\top] $,
  where for $t_1,t_2 \in FX$:
  \begin{align*}
    d^{\uparrow \Gamma}(t_1,t_2) :=
    \sup\{d_e(\gamma(Ff(t_1)),\gamma(Ff(t_2))) \mid f\colon (X,d) \neto
    ([0,\top], d_e), \gamma\in\Gamma \}
  \end{align*}
  We define $\bar{F}_\Gamma(X,d) = (FX,d^{\uparrow \Gamma})$ on
  objects, while $\bar{F}_\Gamma$ is the identity on arrows.
\end{defi}

We will abbreviate $\tilde{F}_\gamma f = \gamma\circ Ff$. Note that
$\tilde{F}_\gamma$ is a functor on the slice category $\Set/[0,\top]$,
which lifts real-valued predicates $p\colon X\to[0,\top]$ to
real-valued predicates $\tilde{F}_\gamma p\colon FX\to[0,\top]$.

It still has to be shown that $\bar{F}$ is well-defined. The proofs
are a straighforward adaptation of the proofs in \cite{BBKK15b}.

\begin{restatable}{lem}{LemLiftingDirectedMetric}
  \label{lem:lifting-directed-metric}
  The Kantorovich lifting for pseudometrics
  (Definition~\ref{def:kantorovich-lifting}) is well-defined, in
  particular it preserves pseudometrics and maps non-expansive
  functions to non-expansive functions.
\end{restatable}

As a sanity check we observe that all evaluation maps
$\gamma\in \Gamma$ are non-expansive for the Kantorovich lifting of
$d_e$\full{ (see Lemma~\ref{lem:nonexp-properties-lifting} in the
  appendix)}. In fact, the Kantorovich lifting is the least lifting
that makes the evaluation maps non-expansive. This also means that a
non-expansive function $f\colon (X,d)\neto([0,1],d_e)$ is always
mapped to a non-expansive
$\tilde{F}_\gamma f\colon (FX,d^{\uparrow \Gamma})\neto ([0,1],d_e)$.

For the following definitions we need the supremum metric on
functions.

\begin{defi}[Supremum metric]
  \label{def:supremum-metric}
  Let $(Y,d)$ be a pseudometric space. Then the set of all functions
  $f\colon X\to Y$ is equipped with a pseudometric $d^\infty$, the
  supremum metric, defined as
  $d^\infty(f,g)= \sup_{x \in X} d(f(x),g(x))$ for $f,g\colon X\to Y$.
\end{defi}

We consider the following restrictions on evaluation maps respectively
predicate liftings, which are needed in order to prove the results.

\begin{defi}[Properties of evaluation maps]
  \label{def:properties-ev-fct}
  Let $\gamma\colon F[0,\top]\to [0,\top]$ be an evaluation map
  for a functor $F\colon Set \to Set$. 
  \begin{itemize}
  \item The predicate lifting $\tilde{F}_\gamma$ induced by $\gamma$
    is non-expansive wrt.\ the supremum metric whenever
    $d_e^\infty(\tilde{F}_\gamma f,\tilde{F}_\gamma g)\le
    d_e^\infty(f,g)$ for all $f,g\colon X\to [0,\top]$ and the same
    holds if we replace $d_e$ by $d_\ominus$.
  \item The predicate lifting $\tilde{F}_\gamma$ is contractive wrt.\
    the supremum metric whenever
    $d_e^\infty(\tilde{F}_\gamma f,\tilde{F}_\gamma g)\le c\cdot
    d_e^\infty(f,g)$ for some $c$ with $0 < c < 1$.
  \item The predicate lifting $\tilde{F}_\gamma$ is
    $\omega$-continuous, whenever for an ascending chain of functions
    $f_i$ (with $f_i \le f_{i+1}$) we have that
    $\tilde{F}_\gamma (\sup_{i<\omega} f_i) =
    \sup_{i<\omega}(\tilde{F}_\gamma f_i)$.
  \end{itemize}
\end{defi}

It can be shown that the first property is equivalent to a property of
the lifted functor, called local non-expansiveness, studied in
\cite{tr:final-metric}.

\begin{restatable}[Local non-expansiveness]{prop}{PropLocalNonexpansive}
  \label{prop:local-nonexpansive}
  Let $\Gamma$ be a set of evaluation maps and let $\bar{F}$ be the
  Kantorovich lifting of a functor $F$ via $\Gamma$. It holds that
  \[ (d_Y^F)^\infty(\bar{F}f,\bar{F}g) \le (d_Y)^\infty(f,g) \] for
  all non-expansive functions $f,g\colon (X,d_X)\to (Y,d_Y)$ (where
  $\bar{F}(Y,d_Y) = (FY,d_Y^F)$) if and only if
  \[ d_e^\infty(\tilde{F}_\gamma f,\tilde{F}_\gamma g) \le
    d_e^\infty(f,g) \] for all non-expansive functions
  $f,g\colon (X,d_X)\to ([0,\top],d_e)$ and all $\gamma\in\Gamma$.
\end{restatable}

\noindent\emph{\textbf{Assumption:} In the following we will always
  assume the first property in Definition~\ref{def:properties-ev-fct}
  for every evaluation map $\gamma$, i.e., the predicate lifting
  $\tilde{F}_\gamma$ is non-expansive wrt.\ the supremum
  metric.}

\smallskip

Under this assumption it can be shown that the Kantorovich lifting
itself is non-expansive (respectively contractive).

\begin{restatable}{prop}{PropLiftingPlusDelta}
  \label{prop:lifting-nonexpansive}
  Let $\Gamma$ be a set of evaluation maps and let
  $d_1,d_2\colon X\times X\to[0,\top]$ be two pseudometrics. Then
  $d_e^\infty(d_1^{\uparrow \Gamma},d_2^{\uparrow \Gamma}) \le
  d_e^\infty(d_1,d_2)$, that is, the Kantorovich lifting of metrics is
  non-expansive for the supremum metric.
  
  If, in addition, every predicate lifting $\tilde{F}_\gamma$ for
  $\gamma\in\Gamma$ is contractive
  (cf. Definition~\ref{def:properties-ev-fct}), we have that
  $d_e^\infty(d_1^{\uparrow \Gamma},d_2^{\uparrow \Gamma}) \le c\cdot
  d_e^\infty(d_1,d_2)$ for some $c$ with $0<c<1$, that is, the
  Kantorovich lifting of metrics is contractive.
\end{restatable}

We will now see that for the functors studied in this paper, we have
evaluation maps that satisfy the required conditions. 

\begin{restatable}{prop}{PropEvNonexpansiveSupContinuous}
  \label{prop:ev-nonexpansive-sup-continuous}
  The following evaluation maps induce predicate liftings which are
  non-expansive wrt.\ the supremum metric and $\omega$-continuous.
  \begin{itemize}
  \item The evaluation map $\gamma_\mathcal{P}$ for the (finite or
    general) powerset functor $\mathcal{P}$ with
    $\gamma\colon \mathcal{P}[0,\top]\to [0,\top]$ where
    $\gamma_\mathcal{P}(R) = \sup R$.
  \item The evaluation map $\gamma_\mathcal{D}$ for the (finitely or
    countably supported) probability distribution functor
    $\mathcal{D}$ (for its definition see Example~\ref{ex:prob-ts})
    with $\gamma_\mathcal{D}\colon \mathcal{D}[0,1]\to [0,1]$ where
    $\gamma_\mathcal{D}(p) = \sum_{r\in [0,1]} r\cdot p(r)$. Note that
    $\gamma_\mathcal{D}$ corresponds to the expectation of the
    identity random variable.
  \item The evaluation map $\gamma_\mathcal{M}$ for the constant
    functor $\mathcal{M}X = [0,\top]$ with
    $\gamma_\mathcal{M}\colon [0,\top]\to [0,\top]$ and
    $\gamma_\mathcal{M}(r) = r$.
  \item The evaluation map $\gamma_\mathcal{S}$ for the constant
    functor $\mathcal{S}X = 1 = \{\bullet\}$ with
    $\gamma_\mathcal{S}\colon 1\to [0,\top]$ and
    $\gamma_\mathcal{S}(r) = \top$.
  \end{itemize}
\end{restatable}

\smallskip

As shown in \cite{BBKK15b} the evaluation map $\gamma_\mathcal{P}$
induces the Hausdorff lifting\footnote{Given a metric $d$ on $X$, the
  Hausdorff lifting of $d$ is the metric $d^H$ with
  $d^H(X_1,X_2) = \max \{\sup_{x_1\in X_1} \inf_{x_2\in X_2}
  d(x_1,x_2), \sup_{x_2\in X_2} \inf_{x_1\in X_1} d(x_1,x_2)\}$ for
  $X_1,X_2\subseteq X$.} on metrics and the evaluation map
$\gamma_\mathcal{D}$ the classical Kantorovich lifting\footnote{Given
  a metric $d$ on $X$, the (probabilistic) Kantorovich lifting of $d$
  is the metric $d^K$ with
  $d^K(p_1,p_2) = \sup\{ |\sum_{x\in X} f(x)\cdot (p_1(x)-p_2(x))|
  \mid f\colon (X,d)\neto ([0,1],d_e)\}$ where
  $p_1,p_2\colon X\to [0,1]$ are probability distributions.} for
probability distributions \cite{v:optimal-transport}.

Contractivity can be typically obtained by using a predicate
lifting which is non-expansive and multiplying with a discount factor
$0 < c < 1$, for instance by using
$\gamma_\mathcal{P}(R) = c\cdot \sup R$ in the first item of
Proposition~\ref{prop:ev-nonexpansive-sup-continuous} above.

It can be shown that the properties of evaluation maps are preserved
under various forms of composition.

\begin{restatable}[Composition of evaluation
  maps]{prop}{PropCompEvaluationMaps}
  \label{prop:comp-evaluation-maps}
  The following constructions on evaluation maps preserve
  non-expansiveness for the supremum metric and $\omega$-continuity
  for the induced predicate liftings.  Let
  $\gamma_F\colon F[0,\top]\to[0,\top]$,
  $\gamma_G\colon G[0,\top]\to [0,\top]$ be evaluation maps for
  functors $F,G$.
  \begin{itemize}
  \item $\gamma\colon F[0,\top]\times G[0,\top]\to[0,\top]$
    with $\gamma = \gamma_F\circ\pi_1$, as an evaluation map for
    $F\times G$.
  \item $\gamma\colon F[0,\top]+G[0,\top]\to[0,\top]$ with
    $\gamma(t) = \gamma_F(t)$ if $t\in F[0,\top]$ and $\gamma(t) = 0$
    otherwise, as an evaluation map for $F+G$.
  \item $\gamma\colon FG[0,\top]\to [0,\top]$ with
    $\gamma = \gamma_F \circ F\gamma_G $, as an evaluation map for
    $FG$.
  \end{itemize}
\end{restatable}

Now we can define behavioural distance on a coalgebra, using the
Kantorovich lifting. Note that the behavioural distance is
  parameterized over $\Gamma$, since, if we are given a coalgebra in
  $\Set$, the notion of behaviour in the metric case is dependent on
  the chosen functor lifting.

\begin{defi}[Behavioural distance]
  \label{def:behavioural-distance}
  Let the coalgebra $\alpha\colon X\to FX$ and a set of evaluation
  maps $\Gamma$ for $F$ be given.  We define the pseudometric
  $d_{\alpha}\colon X \times X \rightarrow [0,\top]$ as the smallest
  fixpoint of
  $d_{\alpha} = d_{\alpha}^{\uparrow \Gamma} \circ (\alpha \times
  \alpha)$.
\end{defi}

Note that every lifting of metrics is necessarily monotone (since it
turns the identity into a non-expansive function,
cf. \cite{BBKK15b}). Since in addition the space of pseudometrics
forms a complete lattice (where $\sup$ is taken pointwise), the
smallest fixpoint exists by Knaster-Tarski.

It has been shown in \cite{bbkk:coalgebraic-behavioral-metrics} that
whenever the Kantorovich lifting preserves metrics (which is the case
for our examples) and the final chain converges, then $d_\alpha$
characterizes behavioural equivalence, i.e., $d_\alpha(x,y) = 0$ iff
$x\sim y$.

\begin{ex}
  \label{ex:prob-metric-ts}
  Using the building blocks introduced above we consider the following
  coalgebras with their corresponding behavioural metrics,
  generalizing notions from the literature. In both cases we are
  interested in the smallest fixpoint.
  \begin{itemize}
  \item \emph{Metric transition systems
      \cite{deAlfaro:2009:LBS:1525651.1525793}:}
    $FX = [0,\top]\times \mathcal{P}X$ with two evaluation maps
    $\gamma_i\colon [0,\top]\times \mathcal{P}[0,\top]\to [0,\top]$,
    $i\in\{1,2\}$ with $\gamma_1(r,R) = r$, $\gamma_2(r,R) = \sup R$.
    
    This gives us the following fixpoint equation, where $d^H$ is the
    Hausdorff lifting of a metric~$d$. Let $\alpha(x) = (r_x,S_x)$,
    $\alpha(y) = (r_y,S_y)$, then
    \[ d(x,y) = \max\{|r_x-r_y|,d^H(S_x,S_y)\} \]
  \item \emph{Probabilistic transition systems:} $GX = \mathcal{D}X+1$
    with two evaluation maps
    $\bar{\gamma}_{\mathcal{D}},\gamma_\bullet\colon
    \mathcal{D}[0,1]+1\to [0,1]$, $i\in\{1,2\}$ with
    $\bar{\gamma}_{\mathcal{D}}(p)= \gamma_\mathcal{D}(p)$,
    $\gamma_\bullet(p) = 0$ where $p\in \mathcal{D}[0,1]$,
    $\bar{\gamma}_{\mathcal{D}}(\bullet) = 0$,
    $\gamma_\bullet(\bullet) = 1$.
    
    This gives us the following fixpoint equation, where $d^K$ is the
    (probabilistic) Kantorovich lifting of a metric $d$. Let
    $T = \{x\mid \alpha(x) = \bullet\}$ and let
    $p_x = \alpha(x) \neq \bullet$.  
    \[ d(x,y) = \left\{
        \begin{array}{ll}
          1 & \mbox{if $x\in T, y\notin T$ or $x\notin T, y\in T$} \\
          0 & \mbox{if $x,y\in T$} \\
          d^K(p_x,p_y) & \mbox{otherwise}
        \end{array}
      \right. \]
  \end{itemize}
\end{ex}

Some of the results on (real-valued) modal logics in
Section~\ref{sec:logics-metric} will require that the fixpoint
iteration terminates in $\omega$ steps. This is related to the fact
that the original Hennessy-Milner theorem requires finite branching.

\begin{restatable}{prop}{PropFixpointOmegaSteps}
  \label{prop:fixpoint-omega-steps}
  Let $\Gamma$ be a set of evaluation maps and let
  $\alpha\colon X\to FX$ be a coalgebra. We define an ascending
  sequence of metrics $d_i\colon X\times X\to [0,\top]$ as follows:
  $d_0$ is the constant $0$-function and
  $d_{i+1} = d_i^{\uparrow \Gamma} \circ \alpha \times
  \alpha$. Furthermore $d_\omega = \sup_{i<\omega} d_i$.
  \begin{itemize}
  \item If for all evaluation maps $\gamma\in\Gamma$ the induced
    predicate liftings are $\omega$-continuous (see
    Definition~\ref{def:properties-ev-fct}) and $F$ is
    $\omega$-accessible, the fixpoint $d_\alpha$ equals $d_\omega$.
  \item If for all evaluation maps $\gamma\in\Gamma$ the induced
    predicate liftings are contractive wrt.\ the supremum metric (see
    Definition~\ref{def:properties-ev-fct}), the fixpoint $d_\alpha$
    equals $d_\omega$.
  \end{itemize}
\end{restatable}

Hence, if we are working with the finite powerset functor or the
finitely supported distribution functor, the first case applies,
whereas in the case of contractiveness, these restrictions are
unnecessary (compare this with the result of \cite{tr:final-metric}
which guarantees the existence of a final coalgebra for a class of
locally contractive functors).

\subsection{Modal Logics  for the Metric Case}
\label{sec:logics-metric}

We now define a coalgebraic modal logic $\mathcal{M}(\Gamma)$, which
is inspired by \cite{VANBREUGEL2005115}. Assume also that $\Gamma$ is
a set of evaluation maps.

The formulas of the logic are defined together with their semantics
$\llbracket \phi\rrbracket_\alpha $ and their modal depth
$\mathit{md}(\phi)$ in Table~\ref{tab:modal_logic}. Given a coalgebra
$\alpha\colon X\to FX$ and a formula $\phi$, the semantics of such a
formula is given by a real-valued predicate
$\llbracket \phi\rrbracket_\alpha\colon X\to [0,\top]$, defined
inductively, where $\gamma\in\Gamma$, $q \in \mathbb{Q}\cap
[0,\top]$. Again we will occasionally omit the subscript $\alpha$.

\begin{center}
  \begin{tabular}{ |l||c|c|c|c|c| } 
    \hline
    $\phi$:&	$\top$ & $[\gamma]\psi$ & 	 $\min(\psi,\psi')$& $\lnot\psi$ & $\psi \ominus q  $ \\	\hline
    $\llbracket \phi\rrbracket_\alpha$:&$\top$ &$\gamma \circ 
    F\llbracket \psi \rrbracket_\alpha \circ \alpha$ &$\min\{\llbracket
    \psi \rrbracket_\alpha, \llbracket \psi' \rrbracket_\alpha\} $ & $\top-\llbracket \psi \rrbracket_\alpha$ & $\llbracket \psi
    \rrbracket_\alpha \ominus q$ \\ 	\hline
    $\mathit{md}(\phi)$:	&	$0$ &$\mathit{md}(\psi)+1$ &$\max\{\mathit{md}(\psi),\mathit{md}(\psi')\}$ & $\mathit{md}(\psi)$ & $\mathit{md}(\psi)$\\ 
    \hline
  \end{tabular}
  \captionof{table}{Overview of the modal logic formulas, their semantics 	 $\llbracket \phi\rrbracket_\alpha$ and modal depths $\mathit{md}(\phi)$.}
  \label{tab:modal_logic}
\end{center}

Note that, given a state $x$ and a logical formula $\phi$, we do
  not just obtain a true value (true, false) dependent on whether $x$
  satisfies the formula or not. Instead we obtain a value in the
  interval $[0,1]$ that measures the degree or weight according to
  which $x$ satisfies $\phi$.

\begin{defi}[Logical distance]
  Let $\alpha\colon X\to FX$ be a coalgebra and let $x,y\in X$. We
  define the logical distance of $x,y$ as
  \[ d_\alpha^L(x,y)= \sup \{d_e(\llbracket \phi
    \rrbracket_\alpha(x), \llbracket \phi \rrbracket_\alpha(y))
    \mid \phi\in\mathcal{M}(\Gamma) \}. \] We also define the logical
  distance up to modal depth~$i$.
  \[ d_i^L(x,y)= \sup \{d_e(\llbracket \phi
    \rrbracket_\alpha(x), \llbracket \phi \rrbracket_\alpha(y))
    \mid \phi\in\mathcal{M}(\Gamma), \mathit{md}(\phi)\le i \}. \]
\end{defi}

\begin{ex}
  \label{ex:metric-formula}
  We are considering probabilistic transition systems with evaluation
  maps as defined in Example~\ref{ex:prob-metric-ts}.
  
  The formula
  $\varphi=[\bar{\gamma}_{\mathcal{D}}][\gamma_\bullet]\top$
  distinguishes the states $x,y$ in Figure~\ref{fig:prob-ts}.  The
  formula $\psi = [\gamma_\bullet]\top$ evaluates to a predicate
  $\llbracket\psi\rrbracket$ that assigns $1$ to terminating states
  and $0$ to non-terminating states. Now $x$ makes a transition to a
  terminating state with probability $\frac{1}{2}$, which means that
  $\llbracket \phi\rrbracket(x) =
  \bar{\gamma}_{\mathcal{D}}(\mathcal{D}\llbracket\psi\rrbracket(\alpha(x)))
  = \frac{1}{2}$. Similarly
  $\llbracket \phi\rrbracket(y) = \frac{1}{2}+\epsilon$.  Hence
  $d^L_\alpha(x,y) \ge d_e(\llbracket\varphi\rrbracket(x),
  \llbracket\varphi\rrbracket(y))=\epsilon$. (In fact, the logical
  distance equals $\epsilon$.)
\end{ex}

We will now show that the logical distance $d_\alpha^L$ and the
  behavioural distance $d_\alpha$ coincide, i.e. a quantitative
  version of the Hennessy-Milner theorem, by generalizing the proof
  from~\cite{VANBREUGEL2005115}. Note that in some respects we
simplify wrt.~\cite{VANBREUGEL2005115} by not working in
$\mathbf{Meas}$, the category of measurable spaces, but in a discrete
setting. On the other hand, we generalize by considering arbitrary
$\Set$-endofunctors. 

\begin{restatable}{theo}{TheoHmRVLogicSound}
  \label{prop:hm-rv-logic-sound}
  Let $d_i$ be the sequence of pseudometrics from
  Proposition~\ref{prop:fixpoint-omega-steps}. Then:
  \begin{enumerate}
  \item \label{prop:hrls-1} For every $i\in\mathbb{N}_0$
    $d_i^L\le d_i$.
  \item \label{prop:hrls-2} For every $\phi$ with
    $\mathit{md}(\phi)\le i$ we have non-expansiveness:
    $\llbracket\phi\rrbracket\colon (X,d_i)\to ([0,\top],d_e)$.
  \item \label{prop:hrls-3} $d_\alpha^L\le d_{\alpha}$.
  \end{enumerate}
\end{restatable}

Note that from Theorem~\ref{prop:hm-rv-logic-sound} it also follows
that for every formula $\phi$ the function $\llbracket\phi\rrbracket$
is non-expansive. Non-expansiveness is analogous to
bisimulation-invariance that holds for formulas in a classical
logic. In particular, in the classical case if $x\sim y$, then
$\llbracket \phi\rrbracket(x) = \llbracket \phi\rrbracket(y)$ for
every $\phi$, in other words $\llbracket\phi \rrbracket$ is
non-expansive for the discrete metric $d$.

The other inequality ($d_\alpha^L\ge d_{\alpha}$) is more difficult to
prove: we will first show that each $d_i$ is totally bounded and then
show that each non-expansive function can be approximated at each pair
of points by a modal formula. Since modal formulas are closed under
$\min$ and $\max$, this enables us to use a variant of a lemma from
\cite{a:real-analysis-probability} to prove that the formulas form a
dense subset of all non-expansive functions. In order to achieve the
approximation, we need all operators of the logic.

We first have to recall some definitions from real-valued analysis.

\begin{defi}[Total boundedness]
  \label{def:total-boundedness}
  A pseudometric space $(X,d)$ is totally bounded iff for every
  $\epsilon > 0$ there exist finitely many elements
  $x_1,\dots,x_n \in X$ such that
  $X = \bigcup_{i=1}^n \mathcal{B}_\epsilon(x_i)$ where
  $\mathcal{B}_\epsilon(x_i) = \{z\in X\mid d(z,x_i)\le \epsilon\}$
  denotes the $\epsilon$-ball around $x_i$.
\end{defi}

Our first result is to show that the lifting preserves total
boundedness, by adapting a proof from \cite{wspk:van-benthem-fuzzy}
from a specific functor to arbitrary functors.

\begin{restatable}{prop}{PropPreservTotalBounded}
  \label{prop:preserv-total-bounded}
  Let $(X,d)$ be a totally bounded pseudometric space, then
  $(FX,d^{\uparrow \Gamma})$ is totally bounded as well.
\end{restatable}

Using this result it can be shown that every pseudometric in the
ascending chain from Proposition~\ref{prop:fixpoint-omega-steps}
(apart from $d_\omega$) is totally bounded.

\begin{restatable}{prop}{PropApproxMetricsTotallyBounded}
  \label{prop:approx-metrics-totally-bounded}
  Let $d_i$ be the sequence of pseudometrics from
  Proposition~\ref{prop:fixpoint-omega-steps}. Then every $(X,d_i)$ is
  a totally bounded pseudometric space.
\end{restatable}

Since total boundedness is not preserved by taking a supremum,
$d_\omega$ is not necessarily totally bounded and we can not iterate
the argument. This is one of the reasons for requiring that the
fixpoint is reached in $\omega$ steps in
Theorem~\ref{thm:hm-rv-logic-expressive} below.

In the next step we show that the formulas are dense in the
non-expansive functions.

\begin{restatable}{prop}{PropFormulasDenseNonexpansive}
  \label{prop:formulas-dense-nonexpansive}
  $\{\llbracket \phi \rrbracket \colon X \rightarrow [0,1] \mid
  \mathit{md}(\phi) \leq i \}$ is dense (wrt.\ the supremum metric)
  in $\{f \colon (X,d_i^L) \neto ([0,\top],d_e)\}$.
\end{restatable}

Finally we can show under which conditions the inequality
$d_\alpha \le d_\alpha^L$ holds.

\begin{restatable}{theo}{ThmHmRVLogicExpressive}
  \label{thm:hm-rv-logic-expressive}
  If the fixpoint $d_\alpha$ is reached in $\omega$ steps, it holds
  that $d_\alpha \le d_\alpha^L$.
\end{restatable}

\subsection{Games for the Metric Case}
\label{sec:games-metric}

We will now present the two-player game characterizing the behavioural
distance between two states. The roles of S and~D are similar to those
in the first game, where D wants to defend the statement that the
distance of two states $x,y \in X $ in a coalgebra $\alpha$ is bounded
by $\epsilon \in [0,\top]$, i.e., $d_\alpha(x,y)\le \epsilon$. S wants
to disprove this claim.

\begin{itemize}
\item \textbf{Initial situation:} Given a coalgebra
  $ \alpha \colon X \to FX $, we start with $(x,y,\epsilon)$ where
  $x,y \in X$ and $\epsilon \in [0,\top]$.
\item \textbf{Step~1:} S chooses $ s \in \{x,y\} $ and a real-valued
  predicate $ p_{1} \colon X \to [0,\top]$.
\item \textbf{Step~2:} D takes $t\in\{x,y\}\backslash\{s\}$ and has to
  answer with a predicate $ p_{2} \colon X \to [0,\top] $, which
  satisfies
  $d_\ominus(\tilde{F}_\gamma p_1(\alpha(s)),\tilde{F}_\gamma
  p_2(\alpha(t))) \leq \epsilon $ for all $\gamma\in \Gamma$.
\item \textbf{Step~3:} S chooses $p_{i} $ with $i \in \{1,2\}$
  and some state $ x' \in X$.
\item \textbf{Step~4:} D chooses some state $ y' \in X$ with
  $ p_i(x') \leq p_j(y') $ where $j\neq i$
\item \textbf{Next round:} $(x',y',\epsilon')$ with
  $\epsilon' = p_j(y')-p_i(x')$.
\end{itemize}

After one round the game continues with the initial step, but now D
tries to show that $d_\alpha(x',y') \leq \epsilon'$. D wins if the
game continues forever. In the other case, e.g., D has no move at
Step~2 or Step 4, S wins.

The game distance of two states is defined as follows.

\begin{defi}[Game distance]
  \label{def:game-distance}
  Let $\alpha\colon X\to FX$ be a coalgebra and let $x,y\in X$. We
  define the game distance of $x,y$ as
  \[ d_\alpha^G(x,y) =\inf \{\epsilon \mid \text{D has a winning
      strategy for } (x,y,\epsilon) \}. \]
\end{defi}

We now prove that the behavioural distance and the game distance
coincide. We first show that $d_\alpha^G$ is indeed a pseudometric.

\begin{restatable}{prop}{PropGameDistancePseudometric}
  \label{prop:game-distance-pseudometric}
  The game distance $d_\alpha^G$ is a pseudometric.
\end{restatable}

Next we show that the game distance is always bounded by the
behavioural distance.

\begin{restatable}{theo}{ThmWinningStrategyDefenderMetric}
  \label{thm:winning-strategy-defender-metric}
  It holds that $d_\alpha^G \le d_\alpha$.
\end{restatable}
\full{While the general proof of this theorem is given in the
  appendix, the strategy for~D can be straightforwardly explained
  whenever $X$ is finite.}\short{While the general proof of this
  theorem is given in \cite{km:bisim-games-logics-metric-arxiv}, the
  strategy for~D can be straightforwardly explained whenever $X$ is
  finite.} In particular we want to show that whenever
$d_\alpha(x,y)\le \epsilon$, then D has a winnning strategy for
$(x,y,\epsilon)$. Assume that $S$ chooses $s \in \{x,y\}$ with
$p_1 \colon X \rightarrow [0,\top]$. In this case D chooses $p_2$ with
$p_2(z) = \sup \{p_1(u)-d_\alpha(u,z) \mid u \in X\}$ in Step~2. From
Lemma~\ref{lem:approx-nonexpansive} we know that $p_1\le p_2$ and
$p_2$ is non-expansive. It can be shown that this choice satisfies
$d_\ominus(\tilde{F}_\gamma p_1(\alpha(s)), \tilde{F}_\gamma
p_2(\alpha(t))) \leq \epsilon $ for all $\gamma\in\Gamma$. Now S
chooses $i$ and $x' \in X$ in Step~3. Then either $i=1$ and D can
choose $y'=x'$ in Step~4 and the game continues with $x',y'$ and
$\epsilon'=p_2(y')-p_1(x')\ge 0$. Or $i=2$ and D can choose $y'$ such
that $p_1(y')-d_\alpha(y',x') = p_2(x')$ (the supremum is reached
since $X$ is finite). This means that $p_1(y')\ge p_2(x')$ and
$\epsilon'=p_1(y')-p_2(x')=d_\alpha(x',y')$. In both cases, the game
can continue.

\begin{ex}
  \label{ex:metric-winning-strategy-duplicator}
  Imagine the initial game situation $(x,y,\epsilon)$ for our example
  in Figure~\ref{fig:prob-ts} and S chooses $x$ with predicate
  $p_1(4) = 1$ and zero for all remaining states. Now D follows the
  strategy above and plays a predicate $p_2$ with
  $p_2(4)=p_2(5)=p_2(7)=1$ and zero for all other states. Since $5,7$
  are at distance~$0$ to $4$, they are now mapped to $1$ as
  well. Since in particular $4$ and $7$ are mapped to $1$, we obtain
  $ d_\ominus(\tilde{\mathcal{D}}_{\bar{\gamma}_{\mathcal{D}}}
  p_1(\alpha(x)), \tilde{\mathcal{D}}_{\bar{\gamma}_{\mathcal{D}}}
  p_2(\alpha(y))) = d_\ominus(\frac{1}{4},\frac{1}{2}+\epsilon)=0 \le
  \epsilon$ (we obtain the same value for $\gamma_\bullet$). Note
  again that D must be allowed to do ``more'' than S. Now the winning
  strategy for D is obvious: if S picks a terminating state $x'$ and
  $p_i$, D can also pick a terminating state $y'$ and $p_j$ with
  $p_j(y')-p_i(x') = 0$ (similarly for non-terminating states). We
  then end up in $(x',y',0)$ where $x',y'$ are behaviourally
  equivalent.
  
  If S had instead chosen $y$ a prediate $p_1$ with $p_1(7) = 1$ and
  zero for all other states, D would choose the same predicate $p_2$
  with
  $ d_\ominus(\tilde{\mathcal{D}}_{\bar{\gamma}_{\mathcal{D}}} p_1(\alpha(y)),
  \tilde{\mathcal{D}}_{\bar{\gamma}_{\mathcal{D}}} p_2(\alpha(x))) =
  d_\ominus(\frac{1}{2}+\epsilon,\frac{1}{2})=\epsilon$.
\end{ex}
We now demonstrate that in the case of infinite branching, the
construction of the winning strategy for the D is not as simple as
described before.

\begin{ex}
  \label{ex:infinite_branching}
  Consider the coalgebra $\alpha\colon X\to \mathcal{D}X + 1$ in
  Figure~\ref{fig:infinite_branching} on the state space
  $X = \{y,y_0,x,x_1,x_2,\dots\}$, where the probability of going from
  $x$ to $x_i$ is $\alpha(x)(x_i) = \frac{1}{2^i}$.
  
  For both states $x,y$ the probability to terminate is $1$ and hence
  $x\sim y$. Now imagine that S selects $x$ and the real-valued
  predicate $p_1$ with $p_1(x_i)=1-\frac{1}{2^i}$ and $p_1(x)=0$. If we would construct the predicate for D as above, via
  $p_2(z)= \sup\{p_1(u)-d_\alpha(u,z) \mid u \in X\}$, this would
  yield $p_2(y_0)=1$ since the distance of all terminating states is
  $0$. 
  {\makeatletter
    \let\par\@@par
    \par\parshape0
    \everypar{}\begin{wrapfigure}{r}{0.4\textwidth}\centering
      \begin{tikzpicture}
        \node (S1) at (0,-2) [circle,draw]{$x$}; 			
        \node [accepting](S4) at (-1.8,-3.5) [circle,draw]{$x_1$};  
        \node [accepting](S5) at (0,-3.5) [circle,draw]{$x_i$}; 
        
        \draw  [->] (S1) to node [left]{$\frac{1}{2^i}$} (S5);	
        \draw  [->] (S1) to node [left ] {$\frac{1}{2^1}$} (S4);		
        \node (...) at (-0.9,-3.6) { $\dots$};
        \node (...) at (0.9,-3.6) { $\dots$};
        \node (y) at (2.5,-2) [circle,draw]{$y$}; 			
        \node [accepting](y2) at (2.5,-3.5) [circle,draw]{$y_0$}; 
        \draw  [->] (y) to node [left]{$1$} (y2);
      \end{tikzpicture}	
      \caption{Probabilistic transition system for the functor
        $FX = \mathcal{D}X+1$, where $X$ is infinite.}
      \label{fig:infinite_branching}
    \end{wrapfigure}
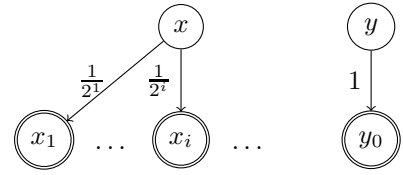Then S chooses $x'=y_0$ and $p_2$ in Step~3 and D
    has no available state $y'$ with which to answer in Step~4. If
    $y'=x_i$, then $p_1(x_i)=1-\frac{1}{2^i} < 1 = p_2(x')$, otherwise
    $p_1(y') = 0 < 1$.
    
    In fact, D has no winning strategy for $\epsilon = 0$, but we can
    show that there is a winning strategy for every $\epsilon > 0$
    (since D can play a predicate that is below $p_2$, but at distance
    $\epsilon$).  Since the game distance is defined as the infinum over
    all such $\epsilon$'s it still holds that $d_\alpha^G(x,y) = 0$.\par}%
\end{ex}
Finally, we can show the other inequality.

\begin{restatable}{theo}{ThmWinningStrategyDefenderMetricRev}
  \label{thm:winning-strategy-defender-metric-rev}
  It holds that $d_\alpha \le d_\alpha^G$.
\end{restatable}

\subsection{Spoiler Strategy for the Metric Case}
\label{sec:spoiler-metric}

The strategy for S for $(x,y,\epsilon)$ can be derived
from a modal formula $\phi$ with
$d_\ominus(\llbracket \phi\rrbracket(x),\llbracket \phi\rrbracket(y))
> \epsilon$. If
$\epsilon < d_\alpha(x,y) = \sup \{d_e(\llbracket
\phi\rrbracket(x),\llbracket \phi\rrbracket(y)) \mid \phi\}$, such a
formula must exist (since we can use negation to switch $x,y$ if
necessary).
The spoiler strategy is defined over the structure of $\phi$:

\begin{itemize}
\item $\phi = \top$: this case can not occur.
\item $\phi = [\gamma]\psi$: S chooses $x$,
  $p_1 = \llbracket\psi\rrbracket$ at Step~1. After D has chosen $y$,
  $p_2$ at Step~2, we can observe that
  $p_2\not\le \llbracket\psi\rrbracket$ \short{(see proof of
  	Theorem~\ref{thm:winning-strategy-spoiler-metric} in \cite{km:bisim-games-logics-metric-arxiv})}\full{(see proof of
  Theorem~\ref{thm:winning-strategy-spoiler-metric})}. Now in Step~3 S
  chooses $p_2$ and $x'$ such $p_2(x')>\llbracket\psi\rrbracket(x')
  $. Now D needs to choose $y'$ such that
  $\llbracket\psi\rrbracket(y')\ge p_2(x')$ in Step~4 and
  $\epsilon' = \llbracket\psi\rrbracket(y') - p_2(x') <
  \llbracket\psi\rrbracket(y') - \llbracket\psi\rrbracket(x') =
  d_e(\llbracket\psi\rrbracket(x'),\llbracket\psi\rrbracket(y'))$ and
  so the game continues in the situation $(x',y',\epsilon')$ with the
  formula $\psi$.
\item $\phi = \min(\psi,\psi')$: In this case either
  $d_e(\llbracket\psi\rrbracket(x),\llbracket\psi\rrbracket(y)) >
  \epsilon$ or
  $d_e(\llbracket\psi'\rrbracket(x),\llbracket\psi'\rrbracket(y)) >
  \epsilon$ \full{(cf. Lemma~\ref{lem:min-max-ms-de}(\ref{lem:mmmd-3})
    in the appendix) }and S picks $\psi$ or $\psi'$ accordingly.
\item $\phi = \lnot \psi$: In this case S takes $\psi$, since
  $d_e(\llbracket\psi\rrbracket(x),\llbracket\psi\rrbracket(y)) =
  d_e(\llbracket\phi\rrbracket(x),\llbracket\phi\rrbracket(y)) >
  \epsilon$.
\item $\phi = \psi\ominus q$: In this case
  $d_e(\llbracket\psi\rrbracket(x),\llbracket\psi\rrbracket(y)) \ge
  d_e(\llbracket\phi\rrbracket(x),\llbracket\phi\rrbracket(y)) >
  \epsilon$
  \full{(cf. Lemma~\ref{lem:min-max-ms-de}(\ref{lem:mmmd-1})) }and
  hence S takes $\psi$.
\end{itemize}

It can be shown that this strategy is indeed correct.

\begin{restatable}{theo}{ThmWinningStrategySpoilerMetric}
  \label{thm:winning-strategy-spoiler-metric}
  Assume that $\alpha\colon X\to FX$ is a coalgebra.  Let $\phi$ be a
  formula with
  $d_e(\llbracket \phi\rrbracket(x),\llbracket \phi\rrbracket(y)) >
  \epsilon$. Then the spoiler strategy described above is winning for
  $S$ in the situation $(x,y,\epsilon)$.
\end{restatable}

Note that Theorem~\ref{thm:winning-strategy-defender-metric-rev} is
not a direct corollary of this theorem, since here we require that a
formula $\phi$ with
$d_e(\llbracket \phi\rrbracket(x),\llbracket \phi\rrbracket(y)) >
\epsilon$ exists, which is not necessarily true in scenarios where the
fixpoint iteration does not terminate in $\omega$ steps.

\begin{ex}
  \label{ex:metric-winning-strategy-spoiler}
  We will show how S can construct a winning strategy for
  $(x,y,\frac{\epsilon}{2})$ based on the formula
  $\varphi=[\bar{\gamma}_{\mathcal{D}}][\gamma_\bullet]\top$ from
  Example~\ref{ex:metric-formula}. The transition system is shown in
  Figure~\ref{fig:prob-ts}. 
  
  It holds that
  $d_\ominus(\llbracket\varphi\rrbracket(y),\llbracket\varphi\rrbracket(x))
  = \epsilon > \frac{\epsilon}{2}$. S plays $y$ and
  $p_1= \llbracket[\gamma_\bullet]\top \rrbracket $ which, due to the
  definition of $\gamma_\bullet$ equals $1$ on terminating states and zero
  on non-terminating states. Now
  $\bar{\gamma}_{\mathcal{D}}(\mathcal{D}p_1(\alpha(y))) = \frac{1}{2}+\epsilon$, so D
  must play in such a way that
  $\bar{\gamma}_{\mathcal{D}}(\mathcal{D}p_2(\alpha(x))) \ge
  \frac{1}{2}+\frac{\epsilon}{2}$. This can only be achieved by
  setting $p_2(3) = \epsilon$ (or to a larger value). Now S chooses
  $p_2$, $x' = 3$ and D can only take $p_1$ and either $4$, $5$ or $7$
  as $y'$. In each case we obtain
  $\epsilon' = p_1(y') - p_2(x') = 1 - \epsilon < 1 = d_e(0,1) = d_e(
  \llbracket[\gamma_\bullet]\top \rrbracket(x'), \llbracket[\gamma_\bullet]\top
  \rrbracket(y'))$.
  
  The spoiler continues to follow his strategy and plays $x'$,
  $p_1 = \llbracket \top\rrbracket$ in the next step, which is
  successful, since $y'$ is a terminating state and $x'$ is not.
\end{ex}

\section{Conclusion}
\label{sec:conclusion}
Comparison to related work can be found in the introduction and
throughout the text.

We will conclude by discussing some open points and questions:
Section~\ref{sec:metric}, which treats the metric cases, follows
  the outline of Section~\ref{sec:classical}, which treats the
  classical case, with some variations. An important difference is the
  fact that the metric case is parameterized over a set $\Gamma$ of
  evaluation maps. Note that we actually mimic the variant of the game
  discussed at the end of Section~\ref{sec:games-classical}, where we
fix evaluation maps, but omit the requirement of weak pullback
preservation. The requirement of monotonicity is replaced by local
non-expansiveness in the metric case. The fact that monotonicity for
partial orders generalizes to non-expansiveness for directed metrics
has already been discussed in \cite{tr:final-metric}.  The variant of the classical game that uses
the lifted order $\le^F$ is more reminiscent of the Wasserstein
lifting for metrics, which has been introduced in \cite{BBKK15b} and
compared to the Kantorovich lifting. It is future work to define a
variant of the metric game that corresponds to the Wasserstein lifting
(or other liftings) of metrics.

Another open question is to prove the Hennessy-Milner theorem for the
real-valued logic in the case where the fixpoint is not reached in
$\omega$ steps. The original variant of the Hennessy-Milner-theorem
only holds for finitely-branching transition systems, but this result
can be generalized if we allow infinite conjunctions (cf. the logic in
Section~\ref{sec:logics-metric}). A natural question is whether the
same solution is applicable to the metric case, by replacing the
$\min$- by an $\inf$-operator (of restricted cardinality $\kappa$, as
in Section~\ref{sec:logics-classical}). However, for this it seems
necessary to generalize the notion of total boundedness to a new
variant where we do not require that the set of ``anchors''
$\{x_1,\dots,x_n\}$ of Definition~\ref{def:total-boundedness} is
finite, but bounded by $\kappa$.

A related question is the following: does the Kantorovich lifting
preserve completeness of metrics? (A metric space $(X,d)$ is complete
if every Cauchy sequence converges in $X$.)
Furthermore we would like to add $\infty$ as a possible distance
value, as in \cite{BBKK15b}. However, this can not be integrated so
easily, for instance it is unclear how to define negation.

Finally, in the quantitative case it could be interesting to know
whether we can use existing efficient algorithms (for the
probabilistic case), for instance in order to generate the strategy of
the spoiler (see e.g. \cite{cbw:complexity-prob-bisimilarity}).
  
\medskip

\bibliography{co_set_game}

\short{\end{document}}


\appendix

\section{Comparison to the Game by Desharnais,
  Laviolette, Tracol}
\label{sec:comparison-desharnais}

As mentioned earlier, the game in \cite{Desharnais} was an inspiration
for our game in the classical case and here we explain the relation in
more detail.

Let $h\colon A\times X\times \mathcal{P}X \to [0,1]$ be a labelled
Markov process, which means that for $a\in A$, $x\in X$,
$S\mapsto h(a,x,S)$ is a sub-probability measure. We simplify for
explanatory purposes and assume that $A$ is a singleton and that
$h(a,x,X)$ is either $1$ or $0$ (either the probabilities sum up to
$1$ or to $0$).  Hence, such a system correspond to a coalgebra
$\alpha\colon X\to \mathcal{D}X + 1$.

In particular, \cite{Desharnais} introduces spoiler-defender games for
$\epsilon$-bisimulation (somewhat similar to our metric games in
Section~\ref{sec:metric}). Here we assume that $\epsilon=0$, which
means that we have exactly probabilistic bisimilarity as defined in
\cite{ls:bisimulation-probabilistic-testing} and in this case the
rules of the game are as follows: consider two states $x,y$.

\begin{itemize}
\item Step~1: S chooses a state $s \in \{x,y\}$. S will play on $s$
  whereas D will play on $t \in \{x,y\} \setminus \{s\}$. S chooses a
  label $a\in A$ and a set $E \subseteq X$.
\item Step~2: D chooses a set $F \subseteq X $, such that
  $h(a,t,F) \geq h(a,s,E)$.
\item Step~3: S chooses a state $x'$ in $E$ or $F$.
\item Step~4: D chooses a state $y'$ in the set not chosen by S. In
  this way we obtain one state in $E$ and one state in $F$.
\end{itemize}

For the functor $FX = \mathcal{D}X+1$ both games are the same: for a
non-terminating state $s$, $Fp_{1}(\alpha(s))$ corresponds to
$ h(a,s,E)$ whenever $E$ is the set specified by $p_1$. Furthermore
$\le^F$ is in this case just the order on the reals (see
Example~\ref{ex:lifted-order}).  If however, $s$ is terminating, i.e.\
$\alpha(s)=\bullet$, we have $Fp_{1}(\alpha(s)) = \bullet$ and
$h(a,s,E) = 0$ for every $E$. Hence both games agree if both states
are terminating or non-terminating.

If however $x$ is terminating and $y$ is not, it is necessary for S in
the game of \cite{Desharnais} to choose $s = y$ and, for instance
$E = X$, since D can not match this move. In our game S can choose
either $x$ or $y$, since $\bullet$ is not related to any real number
via $\le^F$ and the inequality can never be satisfied.


This can also be extended to functors of type $(\mathcal{D}X+1)^L$,
where $L$ represents a finite set of labels or to sub-probability
distributions. For sub-probability distributions, the same phenomenon
as above appears: whenever there are two states whose sub-probability
distributions do not sum up to the same value, S will always win in
Step~2, regardless of his choice, where in the game of
\cite{Desharnais}, S has to pick the state with the larger value.

\section{Proofs}
\label{sec:proofs}

\subsection{Logics and Games for the Classical Case}

\subsubsection{Foundations for the Classical Case}

\LemLiftOrderPred*

\begin{proof}
  We have two predicates $ p_{1} \leq p_{2} $. Define
  $ g \colon X \to \ \leq $ as follows:
  $ g(x) = (p_{1}(x),p_{2}(x)) \in$ $ \leq $, which implies
  $ \pi_{i} \circ g = p_{i} $.
  
  Let $ s \in FX $ and we have to show that $ Fp_1(s) \le^F
  Fp_2(s)$. For this, we require the existence of some
  $ t \in F(\leq) $ with $ F\pi_i (t) = Fp_i(s) $ for
  $i\in\{1,2\}$. Set $ t= Fg(s) $ , then it holds that
  $ F\pi_{i}(t) = F\pi_{i}(Fg(s)) = F(\pi_{i} \circ g) (s) = Fp_{i}(s)
  $.
\end{proof}

\subsubsection{Modal Logics for the Classical Case}

\PropHmLogic*

\begin{proof}~
  
  \noindent\emph{Soundness:} Assume that $x\sim y$, that is there
  exists a coalgebra morphism $f\colon X\to Y$ between $\alpha$ and a
  coalgebra $\beta\colon Y\to FY$ such that
  $\beta\circ f = Ff\circ \alpha$. We show that for all $\phi$,
  $\llbracket \phi \rrbracket_\alpha = \llbracket \phi
  \rrbracket_\beta\circ f$ by structural induction over $\phi$. The
  only interesting case is the modality ($\phi = [\lambda]\psi$) and
  here we obtain
  \[
    \llbracket [\lambda]\psi \rrbracket_{\alpha} = \lambda\circ
    F\llbracket\psi\rrbracket_\alpha\circ \alpha = \lambda\circ
    F(\llbracket\psi\rrbracket_\beta\circ f)\circ \alpha =
    \lambda\circ F\llbracket\psi\rrbracket_\beta\circ
    Ff\circ \alpha = \lambda\circ F\llbracket\psi\rrbracket_\beta\circ
    \beta\circ f = \llbracket [\lambda] \psi \rrbracket_\beta\circ f
  \]
  where the second equality is due to the induction hypothesis.  Now
  it follows easily that
  $\llbracket \phi \rrbracket_\alpha(x) = \llbracket \phi
  \rrbracket_\beta(f(x)) = \llbracket \phi \rrbracket_\beta(f(y)) =
  \llbracket \phi \rrbracket_\alpha(y)$.
  
  \smallskip
  
  \noindent\emph{Expressiveness:} We define a logical
  indistinguishability relation on $X$:
  $x \equiv y \iff \forall \phi (x\models \phi \iff y \models
  \phi)$. Let $f\colon X\to X/ \equiv$ be the function that
  maps every $x\in X$ to its equivalence class $[x]_\equiv$. In
  particular $f(x) = f(y)$.
  
  Furthermore we define
  $\beta\colon (X/ \equiv) \to F(X/ \equiv)$ as
  follows: $\beta([x]_\equiv) = Ff(\alpha(x))$. It is sufficient to
  show that $\beta$ is well-defined, since then
  $\beta\circ f = Ff\circ \alpha$ and we have shown that $x\sim y$.
  
  Hence we have to show that for $x,y\in X$ with $x\equiv y$ it always
  holds that $Ff(\alpha(x)) = Ff(\alpha(y))$. Since $\Lambda$ is
  separating for $F$ (cf. Definition~\ref{def:separating-functor}) it
  is sufficient to show that
  \begin{equation}
    (\lambda\circ Fp\circ Ff\circ \alpha)(x) = (\lambda\circ Fp\circ
    Ff\circ \alpha)(y) \label{eq:sound-complete-1}
  \end{equation}
  for all $\lambda\in\Lambda$ and all
  $p\colon (X/\equiv)\to 2$.
  
  Since $F$ is $\kappa$-accessible there exists $Z\subseteq X$,
  $|Z| < \kappa$ such that $\alpha(x),\alpha(y) \in FZ \subseteq FX$.
  
  Now, for a given $p\colon (X/\equiv)\to 2$ define $q\colon X\to 2$
  with $q = p\circ f$.
  
  Given $z_0,z_1\in Z$ with $q(z_0) = 0$, $q(z_1) = 1$, we know that
  $f(z_0) \neq f(z_1)$ and hence $z_0\not\equiv z_1$. So there exists
  a formula $\phi_{z_0,z_1}$ that distinguishes $z_0,z_1$, that is
  $z_0\not\models \phi_{z_0,z_1}$ and $z_1\models \phi_{z_0,z_1}$
  (this can be ensured since we have negation).
  
  Now consider the following formula:
  \[ \phi = \bigwedge_{\substack{z_0\in Z\\q(z_0)=0}} \big(
    \bigvee_{\substack{z_1\in Z\\q(z_1)=1}} \phi_{z_0,z_1}\big) \]
  Note that since $|Z| < \kappa$ the conjunction and disjunctions obey
  the cardinality restrictions. For every $z_1\in Z$ with $q(z_1) = 1$
  a formula $\phi_{z'_0,z_1}$, which is satisfied by $z_1$, occurs in
  every disjunction and hence $\llbracket \phi\rrbracket(z_1) = 1$. On
  the other hand for every $z_0\in Z$ with $q(z_0) = 0$ there is a
  disjunction consisting only of formulas $\phi_{z_0,z'_1}$, which are
  not satisfied by $z_0$, and hence
  $\llbracket \phi\rrbracket(z_0) = 0$. Summarizing, we obtain that
  $\llbracket \phi\rrbracket$ and $q$ agree on $Z$.
  
  We will now proceed to show that~(\ref{eq:sound-complete-1}) holds
  for a given $\lambda$ and $p$: set $\psi = [\lambda]\phi$ and we
  obtain
  \begin{eqnarray*}
    && (\lambda\circ Fp\circ Ff\circ \alpha)(x) 
    = (\lambda\circ F(p\circ f)\circ \alpha)(x) 
    = (\lambda\circ F(q)\circ \alpha)(x) \\
    & = & (\lambda\circ F\llbracket\phi\rrbracket\circ \alpha)(x) 
    = \llbracket\psi\rrbracket(x) 
  \end{eqnarray*}
  Here we use the fact that whenever $t\in FZ$ and $f|_Z = g|_Z$ for
  $f,g\colon X\to 2$ we have $Ff(t) = Fg(t)$ (since
  $f\circ\iota = g\circ\iota$, where $\iota\colon Z\to X$ is the
  embedding of $Z$ into $X$). In our case $t = \alpha(x)$.
  
  Similarly we have
  $(\lambda\circ Fp\circ Ff\circ \alpha)(x) =
  \llbracket\psi\rrbracket(y)$ and since $x,y$ are logically
  indistinguishable (\ref{eq:sound-complete-1}) follows. 
\end{proof}

\PropSepPlAsJi*

\begin{proof}~ 
  
  \noindent ($\Rightarrow$) It follows directly from
  \cite{SCHRODER2008230} that
  $\{Fp \colon FX \rightarrow F2\}_{p\colon X \rightarrow 2}$ is jointly
  injective.
  
  We now show that $\le^F$ is anti-symmetric.  Let $t_1,t_2 \in F2$
  with $t_1 \leq^{F} t_2 \leq^{F} t_1 $ and $t_1 \neq t_2$. Then there
  exists a monotone $\lambda \in \Lambda $ and some
  $p: 2 \rightarrow 2$ such that
  $\lambda \circ Fp(t_1) \neq \lambda \circ Fp(t_2) $.  We consider
  the following two cases:
  \begin{itemize}
  \item \emph{$p$ is monotone:} In this case $Fp$ is also monotone,
    which can be shown as follows.  Assume that $s_1 \leq^F s_2$, so
    there exists some $s \in F(\le)$, such that $F\pi_i(s) =
    s_i$. Now, since $p$ is monotone, $p\times p$ can be restricted to
    $p\times p\colon \!\le\ \to \ \le$. Furthermore
    $ F\pi_i \circ F(p \times p)(s) = Fp \circ F\pi_i (s)$ for
    $i = 1,2$ and $t'=F(p \times p)(s) \in F(\le)$ is a witness for
    $Fp(s_1) \leq^{F} Fp(s_2) $.
    
    Now, since $\lambda$ is also monotone,
    $t_1 \leq^{F} t_2 \leq^{F} t_1 $ implies
    $\lambda\circ Fp(t_1) \le \lambda\circ Fp(t_2)\le \lambda\circ
    Fp(t_1)$, hence $\lambda \circ Fp(t_1) = \lambda \circ Fp(t_2)$,
    which is a contradiction.
    
  \item \emph{$p$ is not monotone:} since $p:2\to 2$, the only
    non-monotone such predicate is $p(0)=1$, $p(1) = 0$, which is
    antitone. Similar to above we can argue that $Fp$ is antitone as
    well and complete the argument. (Here
    $p\times p\colon \!\le\ \to \ \ge$ and we choose
    $s'=F( \mathit{sym} \circ(p \times p))(s)$, where
    $\mathit{sym} \colon 2\times 2\to 2\times 2$ switches its arguments.)
  \end{itemize}
  
  \noindent ($\Leftarrow$) We assume that $\leq^{F}$ is anti-symmetric
  and that there exists a separating set of evaluation maps, i.e.,
  $\{Fp\colon FX \rightarrow F2\}_{p \colon X \rightarrow2}$ is jointly
  injective. Now let $t_1 \neq t_2$ and $t_1,t_2 \in F2$. Since
  $\{Fp\}_p$ is jointly injective, there exists a predicate such that
  $Fp(t_1) \neq Fp(t_2)$. Now due to the antisymmetry of
  $\le^F$ two cases can occur and we show that in each case there
  exists a monotone evaluation map which separates the two:
  \begin{itemize}
  \item $Fp(t_1) \nleq^{F} Fp(t_2) $: We define
    $\lambda \colon F2 \rightarrow 2$ as follows:
    \begin{equation}
      \lambda(t) =
      \begin{cases}
        0 & \text{if } t \leq^{F} Fp(t_2) \\
        1 & \text{otherwise}
      \end{cases}
    \end{equation}
    Note that $\lambda$ is monotone and that $\lambda(Fp(t_2)) = 0$,
    $\lambda(Fp(t_1)) = 1$.
  \item $Fp(t_2) \nleq^{F} Fp(t_1) $: analogously.
  \end{itemize}
\end{proof}

\begin{prop}
  \label{prop:ev-monotone}
  Let $\lambda\colon FV\to V$ be an evaluation map mapping to an
  ordered set $(V,\le)$. It induces a monotone predicate lifting
  $(p\colon X\to V)\mapsto (\lambda\circ Fp\colon FX\to V)$ iff
  $\lambda\colon (FV,\le^F)\to (V,\le)$ is monotone.
\end{prop}

\begin{proof}~
  
  \noindent ($\Leftarrow$) Assume that the predicate lifting is
  monotone, i.e., given two predicates $p_1\le p_2$ it holds that
  $\lambda\circ Fp_1 \le \lambda\circ Fp_2$. In order to show
  monotonicity of $\lambda$ take $v_1,v_2\in FV$ such that
  $v_1\le^F v_2$.  This means that there exists $r\in F(\le)$ such
  that $F(\pi_1\circ o)(r) = v_1$, $F(\pi_2\circ o)(r) = v_2$, where
  $o\colon \le\ \to V\times V$ is the embedding of the order into
  $V\times V$.
  
  Now consider $\pi_1\circ o$, $\pi_2\circ o\colon \le\ \to V$. It
  holds that $\pi_1\circ o\le \pi_2\circ o$ and with monotonicity of
  the lifting we can conclude
  $\lambda \circ F(\pi_1\circ o) \le \lambda\circ F(\pi_2\circ o)$.
  
  Hence
  $\lambda(v_1) = \lambda(F(\pi_1\circ o)(r)) \le \lambda(F(\pi_2\circ
  o)(r)) = \lambda(v_2)$, i.e., we have shown that $\lambda$ is
  monotone.
  
  \medskip
  
  \noindent ($\Rightarrow$) Assume that $\lambda$ is monotone and take
  $p_1,p_2\colon X\to V$ such that $p_1\le p_2$. 
  Hence $Fp_1(t) \le^F Fp_2(t)$ by Lemma~\ref{lem:lift-order-pred}.
  
  Then we can conclude that $\lambda(Fp_1(t)) \le \lambda(Fp_2(t))$,
  using the monotonicity of $\lambda$.
\end{proof}

\subsubsection{Games for the Classical Case}

\ThmWinningStrategyDefender*

\begin{proof}~
  
  \noindent\emph{$ x \sim y$ $\Rightarrow$ D
    has winning strategy:} We show that whenever $x\sim y$, then
  D can always answer the steps of S and we end up in a
  pair $x'\sim y'$, from which this strategy continues.
  
  Whenever $ x \sim y $, there exists a coalgebra
  $ \beta \colon Z\to FZ $ and a coalgebra homomorphism
  $ f \colon X \to Z $ such that $ f(x) = f(y) $.
  
  We assume that S chooses state~$x$ (the other case is
  analogous) and a predicate $ p_{1} \colon X \to 2$. D has now
  to react with a predicate $ p_{2} $. This is constructed by setting
  $ p_{2}(x)=1$ for $ x \in X $ whenever there exists $ x'\in X$ such
  that $f(x) = f(x')$ and $p_{1}(x')=1 $. In other words, we set
  $p_2 = p'_1\circ f$, where $p'_1\colon Z\to 2$ is the least
  predicate such that $p'_1\circ f\ge p_1$ (i.e., $ p_{1}' \leq p$ for
  all $ p $ satisfying $ p\circ f \geq p_{1} $).
  
  
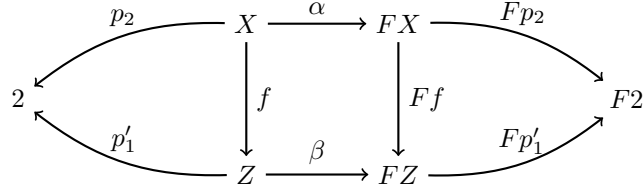
\begin{center}
    \begin{tikzpicture}
      \node (2) at (-3,-1)  {$ 2 $};
      \node (X) at (0,0)  {$ X $};
      \node (FX) at (2,0)  {$ FX $};
      \node (Z) at (0,-2)  {$ Z $};
      \node (FZ) at (2,-2)  {$ FZ $};
      \node (F2) at (5,-1)  {$ F 2 $};
      \draw[->,   thick] (X) to node[right] {$f $} (Z);
      \draw[->,   thick] (X) to node[above] {$\alpha$} (FX);
      \draw[->,   thick] (Z) to node[above] {$\beta$} (FZ);
      \draw[->,   thick] (FX) to node[right] {$Ff$} (FZ);
      \draw[->,   thick] (X) to [bend right=20] node[above] {$p_{2}$} (2);
      \draw[->,   thick] (Z) to [bend left=20] node[above] {$p_{1}'$} (2);
      \draw[->,   thick] (FX) to [bend left=20] node[above] {$Fp_{2}$} (F2);
      \draw[->,   thick] (FZ) to [bend right=20] node[above] {$Fp_{1}'$} (F2);
    \end{tikzpicture}
  \end{center}
  
  Since $ p_{1} \leq p_{2} $, we know by Lemma~\ref{lem:lift-order-pred}
  that $Fp_{1} \leq^F Fp_{2}$ holds.
  
  We obtain:
  \begin{align*}
    && (Ff \circ \alpha)(x) = (\beta \circ f)(x)&= (\beta \circ 
    f)(y)=(Ff \circ \alpha)(y)\\
    \text{ which implies }\\
    &&(Fp_1'\circ Ff \circ \alpha)(x) &= (Fp_1'\circ Ff \circ \alpha)(y) \\
    &&\Rightarrow (Fp_2 \circ \alpha)(x) &= (Fp_2 \circ \alpha)(y)\\
    &&\Rightarrow (Fp_1 \circ \alpha)(x) &\leq^{F} (Fp_2\circ \alpha)(x) 
    = (Fp_2 \circ \alpha)(y)
  \end{align*}
  
  S now chooses a predicate $p_i$ and a state $x'$ with the
  constraints described in Step~3, i.e., $p_i(x') = 1$. (If S
  can not give such a predicate and state, D wins
  automatically.)
  
  If S chooses $p_1$ and $x'$, D can simply pick $y' =
  x'$, since $p_2(y') \ge p_1(x') = 1$. In this case we end up in
  $(x',x')$ with $x'\sim x'$.
  
  If D chooses $p_2$ and $x'$, by construction of $p_2$
  D can find a state $y'$ with $f(x') = f(y')$ and $p_1(y') =
  1$. In this case $x'\sim y'$ holds.
  
  In both cases the game can continue.
  
  \medskip
  
  \noindent \emph{D has a winning strategy
    $ \Rightarrow x \sim y $:} First, we need to show that the
  relation 
  \[ W= \{(x,y) \in X \times X \mid \text{there exists a winning
      strategy of D for } (x,y) \} \] is an
  equivalence.
  
  \begin{itemize}
  \item \emph{$ W $ is reflexive:} $ (x,x) \in W $ for every $x\in X$.
    
    Assume that S chooses $x$ and $p_{1}$, then D chooses $x$ and
    $p_{1}$ as well, for which we clearly have
    $Fp_{1} (\alpha (x) ) \leq^{F} Fp_{1} (\alpha (x) )$.
    
    Then the next game situation is $(x',x')$, for which we can
    continue this strategy forever. 
    
  \item \emph{$ W $ is symmetric:} $ (x,y) \in
    W$ implies $(y,x) \in W $. 
    
    If there is a winning strategy for $(x,y)$ there must always also
    be a winning strategy for $(y,x)$, since S can choose
    either $x$ or $y$.
  \item \emph{$ W $ is transitive:} 
    if $ (x,y), (y,z) \in W $, then $ (x,z) \in W $.
    
    Assume that in Step~1 S chooses $x$ and $p_{1}$ (the case
    where S chooses $y$ is analogous, taking into account that
    $W$ is symmetric).  We know by $(x,y) \in W$ that D has an
    answer, hence he chooses $p_2$, for which
    $Fp_{1}(\alpha (x) ) \leq^{F} Fp_{2} (\alpha (y) )$.
    
    If S were to make the choice of $p_2$ and $y$, we know by
    $(y,z) \in W$ that D has an answering move, by choosing
    $p_2$ such that
    $Fp_{2} (\alpha (y) ) \leq^{F} Fp_{3} (\alpha (z) )$.
    
    Hence D makes the choice of $p_3$ in Step~2.  Now we have
    that
    $Fp_{1} (\alpha (x) ) \leq^{F} Fp_{2} (\alpha (y) ) \leq^{F}
    Fp_{3} (\alpha (z) )$ and, by transitivity,
    $Fp_{1} (\alpha (x) ) \leq^{F} Fp_{3} (\alpha (z) )$. (Note that
    transitivity holds since $F$ preserves weak pullbacks.)
    
    Assume that in Step~3 S chooses $p_1$, $x'$ with
    $p_1(x') = 1$. Again, by $(x,y)\in W$, there is an answer of
    D who chooses $y'$ with $p_2(y') = 1$ and $(x',y')\in
    W$. From $(y,z)\in W$ we know that if S chooses $p_2$,
    $y'$, there is an answer by D who chooses $z'$ with
    $p_3(z') = 1$ and $(y',z')\in W$. This state $z'$ is hence finally
    chosen by D in Step~4. 
    
    If instead in Step~3 S chooses $p_3$, $z'$, the choice
    propagates in the other direction.
    
    Since we now have $(x',y'), (y',z')\in W$, we can continue this
    strategy for D forever.
  \end{itemize}
  
  We define a function $ f \colon X \rightarrow Y $ with $ Y= X/W $
  and $ f(x)= [x]_W $.
  
  \begin{center}
    \begin{tikzpicture}
      \node (X) at (0,0)  {$ X $};
      \node (FX) at (2,0)  {$ FX $};
      \node (Y) at (0,-2)  {$ Y $};
      \node (FY) at (2,-2)  {$ FY $};
      \draw[->,   thick] (X) to node[left] {$f $} (Y);
      \draw[->,   thick] (X) to node[above] {$\alpha$} (FX);
      \draw[->,   thick] (Z) to node[above] {$\beta$} (FY);
      \draw[->,   thick] (FX) to node[right] {$Ff$} (FY);
    \end{tikzpicture}
  \end{center}
  
  It suffices to show that $ \beta(f(x)) := Ff(\alpha(x)) $ is
  well defined, since then we have a coalgebra homomorphism that
  witnesses the behavioural equivalence of $x,y$. (Note that $f(x) =
  f(y)$, since both are contained in the same equivalence class.)
  
  Assume that we have a winning strategy for $ (x,y) $ (i.e.,
  $(x,y)\in W$) or in other words $ f(x)=f(y)$, but
  $ Ff(\alpha(x)) \neq Ff(\alpha(y))$. Then we know, by the assumption
  that the functor $F$ has a separating set of predicate liftings
  (respectively the equivalent condition in \cite{SCHRODER2008230}),
  that some $ p \colon Y \rightarrow 2$ exists such that
  $ Fp(Ff(\alpha(x))) \neq Fp(Ff(\alpha(y))) $.
  
  We now show, by contradiction, that D does not have a winning
  strategy for $(x,y)$: S chooses $ p_1 = p \circ f$ and we obtain
  $ Fp_1 \circ \alpha(x) \neq Fp_1 \circ \alpha(y) $, since
  $ Fp_1 = Fp \circ Ff $. Since the preorder $ \leq^F $ on $ F2 $ is
  antisymmetric due to Proposition~\ref{prop:separating-pl-antisym-ji}
  at least one of the following two overlapping cases will occur:
  
  \begin{itemize}
  \item $Fp_1 \circ \alpha(x) \nleq^{F} Fp_1 \circ \alpha(y)$
  \item $Fp_1 \circ \alpha(y) \nleq^{F} Fp_1 \circ \alpha(x)$
  \end{itemize}
  
  Here we only consider the first case, since for the second case the
  argument is analogous. S picks $ x $ and D can not play $p_2$ such
  that $p_2\le p_1$, since in this case we would get
  $Fp_2 \leq^F Fp_1$. Combining this with the condition of Step~3 we
  obtain
  $ (Fp_1 \circ \alpha)(x) \leq^F (Fp_2 \circ \alpha)(y) \le^F (Fp_1
  \circ \alpha)(y)$ which, with transitivity, is a contradiction to
  the first case above.
  
  Hence $p_2\nleq p_1$, which implies that some $ x' \in X $ exists
  such that $ p_2(x')=1 $ and $ p_1(x') =0$. So S picks $ p_2 $ and
  $ x' $. D then picks some $ y' \in X $ with $ p_1(y')=1 $. If
  $( x', y') \in W $ it follows from the construction of
  $p_1 = p \circ f $ that $ p_1(x') =p_1(y')$, but this is again a
  contradiction to $ p_1(x')=0 $. Hence $(x',y')\not\in W$ and D does
  not have a winning strategy.
\end{proof}

\subsubsection{Spoiler Strategy for the Classical Case}

\ThmWinningStrategySpoiler*

\begin{proof}
  Each step described in the strategy yields a smaller formula by
  structural induction and a pair of states which is distinguished by
  the formula. Hence the game will eventually terminate.
  
  We only have to consider the case $\phi = [\lambda]\psi$ in more
  detail and show (by contradiction) that S can make a valid move in
  Step~3 by proving that the predicate $p_2$ chosen by D in Step~2
  must satisfy $p_2\not\le \llbracket\psi\rrbracket$.
  
  Hence, assume that $p_2 \le \llbracket\psi\rrbracket$.  From
  Lemma~\ref{lem:lift-order-pred} and from the monotonicity of
  $\lambda$ it follows that
  \[ (\lambda\circ Fp_2 \circ\alpha)(y) \le (\lambda\circ
    F\llbracket\psi\rrbracket \circ \alpha)(y) =
    \llbracket\phi\rrbracket(y). \] Since $y\not\models \phi$ the
  right-hand side, as well as the left-hand side of the inequality
  must be $0$. On the other hand
  $Fp_1 \circ \alpha(x) \leq^F Fp_2 \circ \alpha(y)$ and hence, again
  due to monotonicity of $\lambda$, we have
  \[ (\lambda\circ F\llbracket\psi\rrbracket \circ\alpha)(x) \le
    (\lambda\circ Fp_2 \circ \alpha)(y). \] Since $x\models\phi$ the
  left-hand side of the inequality must be $1$. But this is a
  contradiction, because then the right-hand side has to be equal to
  $1$ as well.
  
  Hence $p_2\not\le \llbracket\psi\rrbracket$, which means that there
  is a $y'\in X$ such that $p_2(y')=1$ and
  $\llbracket \psi\rrbracket(y') = 0$, i.e., $y'\not\models
  \psi$. 
\end{proof}

\subsection{Logics and Games for the Metric Case}

\subsubsection{Foundations for the Metric Case}

\LemApproxNonexpansive*

\begin{proof}
  The first obvious fact is that, $f \leq h$ holds, since
  $f(z) = f(z)-d(z,z) \leq \sup \{f(z) -d(u,z) \mid u \in X\}=
  h(z)$. 
  
  Next, we show that $h$ is non-expansive. Let $z,z'\in X$. Due to the
  definition of $h$ there exists for all $\delta > 0$ an $u \in X$
  with $h(z) \le f(u) - d(u,z) + \delta$ and
  $f(u) - d(u,z') \le h(z')$. Combined, we have
  \[ h(z)-h(z') \leq f(u) - d(u,z) + \delta - f(u) + d(u,z') = d(u,z')
    - d(u,z) + \delta. \] Since this holds for every $\delta > 0$ we
  have $h(z)-h(z') \le d(u,z') - d(u,z)$.  Due to the triangle
  inequality $d(u,z') \le d(u,z) + d(z,z')$, hence
  $h(z) - h(z') \le d(u,z') - d(u,z)\le d(z,z')$.
  
  Analogously, we can show $h(z') - h(z) \le d(z',z) = d(z',z)$ (due
  to symmetry), hence $d_e(h(z),h(z')) \le d(z,z')$, which means that
  $h$ is non-expansive.
  
  For the function $g$ the proof is analogous.
\end{proof}

\LemLiftingDirectedMetric*

\begin{proof}
  $\bar{F}$ preserves identities and composition of arrows, since $F$
  does. Hence we only have to show the following.
  
  \noindent\emph{Preservation of pseudometrics:} we show that
  reflexivity and triangle inequality are preserved. Assume that
  $t,t_1,t_2,t_3 \in FX$ and let $d\colon X\times X\to [0,\top]$ be a
  pseudometric.
  \begin{itemize}
  \item \emph{Reflexivity} holds since $d_e$ is reflexive:
    \[ d^{\uparrow \Gamma}(t,t) = \sup\{d_e(\tilde{F}_\gamma
      f(t),\tilde{F}_\gamma f(t)) \mid f: (X,d) \neto ([0,\top], d_e),
      \gamma\in\Gamma \} = 0 \]
  \item \emph{Symmetry} holds since $d_e$ is symmetric:
    \begin{eqnarray*}
      d^{\uparrow \Gamma}(t_1,t_2) & = & \sup\{d_e(\tilde{F}_\gamma
      f(t_1),\tilde{F}_\gamma f(t_2)) \mid f: (X,d) \neto
      ([0,\top], d_e), \gamma\in\Gamma  \} \\
      & =& \sup\{d_e(\tilde{F}_\gamma f(t_2),\tilde{F}_\gamma f(t_1))
      \mid f: (X,d) \neto ([0,\top], d_e), \gamma\in\Gamma \} =
      d^{\uparrow \Gamma}(t_2,t_1)
    \end{eqnarray*}
  \item \emph{Triangle inequality:} since $d_e$ is a pseudometric it
    satisfies the triangle inequality, in particular:
    \[ d_e(\tilde{F}_\gamma f(t_1),\tilde{F}_\gamma f(t_3))\leq
      d_e(\tilde{F}_\gamma f(t_1),\tilde{F}_\gamma f(t_2)) +
      d_e(\tilde{F}_\gamma f(t_2),\tilde{F}_\gamma f(t_3)) \] for all
    $f\colon (X,d)\neto ([0,\top],d_e)$. Therefore we take the
    supremum on both sides and obtain
    \begin{eqnarray*}
      d^{\uparrow \Gamma}(t_1,t_3) & = & 
      \sup_{f,\gamma} d_e(\tilde{F}_\gamma f(t_1),\tilde{F}_\gamma f(t_3)) \\
      & \leq & \sup_{f,\gamma}
      \big(d_e(\tilde{F}_\gamma f(t_1),\tilde{F}_\gamma f(t_2)) + 
      d_e(\tilde{F}_\gamma f(t_2),\tilde{F}_\gamma f(t_3))\big) \\
      & \leq & \sup_{f,\gamma}
      d_e(\tilde{F}_\gamma f(t_1),\tilde{F}_\gamma f(t_2)) + \sup_{f,\gamma}
      d_e(\tilde{F}_\gamma f(t_2),\tilde{F}_\gamma f(t_3))  \\
      & = &
      d^{\uparrow \Gamma}(t_1,t_2) + d^{\uparrow \Gamma}(t_2,t_3).
    \end{eqnarray*}
  \end{itemize}
  
  \noindent\emph{Non-expansive functions:} $\bar{F}_\Gamma$ preserves 
  non-expansive functions: Let $f\colon (X,d_X) \neto (Y,d_Y)$ be
  non-expansive and $t_1, t_2 \in FX$, then
  \begin{eqnarray*}
    d_Y^{\uparrow \Gamma}\big(Ff(t_1), Ff(t_2)\big) & = & \sup_{\gamma\in
      \Gamma} \sup_{g \colon (Y,d_Y) 
      \neto ([0,\top], d_e)} 
    d_e\Big(\tilde{F}_\gamma (g \circ f)(t_1),\tilde{F}_\gamma 
    (g \circ f)(t_2)\Big) \\
    & \leq & \sup_{\gamma\in
      \Gamma} \sup_{h \colon (X,d_X) \neto ([0,\top],d_e)}
    d_e\Big(\tilde{F}_\gamma (h)(t_1),\tilde{F}_\gamma (h)(t_2)\Big)
    = d_X^{\uparrow \Gamma}(t_1, t_2)
  \end{eqnarray*}
  due to the fact that since both $f$ and $g$ are non-expansive, also
  their composition
  $(g \circ f) \colon (X,d_X) \neto ([0,\top],d_e)$ is
  non-expansive.
  
\end{proof}

\PropLocalNonexpansive*

\begin{proof}~
  
  \noindent ``$\Rightarrow$'' Let $t\in FX$. By choosing the
  non-expansive identity function in the Kantorovich lifting we
  obtain
  $d_e^\infty(\tilde{F}_\gamma f(t),\tilde{F}_\gamma g(t)) =
  d_e^\infty(\tilde{F}_\gamma \mathit{id}(Ff(t)),\tilde{F}_\gamma
  \mathit{id}(Fg(t))) \le d_e^{\uparrow \Gamma}(Ff(t),Fg(t)) \le
  d_e^\infty(f,g)$. Note that in this setting $d_Y = d_e$ and
  $d_Y^F = d_e^{\uparrow \Gamma}$.
  
  \smallskip
  
  \noindent ``$\Leftarrow$'' Let again $t\in FX$.
  \begin{eqnarray*}
    && (d_Y^F)^\infty(\bar{F}f(t),\bar{F}g(t)) \\
    & = & (d_Y)^{\uparrow \Gamma}(Ff(t),Fg(t)) \\
    & = & \sup \{d_e(\tilde{F}_\gamma h(Ff(t)),\tilde{F}_\gamma
    h(Fg(t))) \mid h\colon (Y,d_Y)\neto ([0,\top],d_e), \gamma\in\Gamma \} \\
    & = & \sup \{d_e(\tilde{F}_\gamma (h\circ f)(t),\tilde{F}_\gamma
    (h\circ g)(t)) \mid h\colon (Y,d_Y)\neto ([0,\top],d_e),
    \gamma\in\Gamma \}
  \end{eqnarray*}
  We know that
  $d_e(\tilde{F}_\gamma (h\circ f)(t),\tilde{F}_\gamma (h\circ g)(t))
  \le d_e^\infty(h\circ f,g\circ h)$ and it is left to show that
  $d_e^\infty(h\circ f,h\circ g) \le (d_Y)^\infty (f,g)$ for every
  such $h$. So let $x\in X$ and we obtain
  $d_e(h(f(x)),h(g(x))) \le d_Y(f(x),g(x))$ since $h$ is
  non-expansive. Taking the supremum on both sides we obtain the
  desired result.
\end{proof}

\begin{lem}
  \label{lem:min-max-ms-de}
  For all $a,b,a_i,b_i,q\in[0,\top]$ it holds that
  \begin{enumerate}
  \item \label{lem:mmmd-1} $d_e(a \ominus q,b\ominus q)\leq d_e(a,b)$.
  \item \label{lem:mmmd-2}
    $d_e(\sup_{i\in I} a_i,\sup_{i\in I} b_i) \le \sup_{i\in I}
    d_e(a_i,b_i)$.
  \item \label{lem:mmmd-3}
    $d_e(\inf_{i\in I} a_i,\inf_{i\in I} b_i) \le \inf_{i\in I}
    d_e(a_i,b_i)$.
  \item \label{lem:mmmd-4}
    $d_\ominus(\sup_{i\in I} a_i,\sup_{i\in I} b_i) \le \sup_{i\in I}
    d_\ominus(a_i,b_i)$.
  \end{enumerate}
\end{lem}

\PropEvNonexpansiveSupContinuous*

\begin{proof}
  In the following let $f,g\colon X\to[0,\top]$.
  \begin{itemize}
  \item Observe that
    $\tilde{\mathcal{P}}_{\gamma_\mathcal{P}}(f)\colon
    \mathcal{P}(X)\to [0,\top]$ satisfies
    $\tilde{\mathcal{P}}_{\gamma_\mathcal{P}}(f)(Z) = \sup f[Z]$ where
    $Z\subseteq X$.
    
    We have
    \begin{eqnarray*}
      && d_e(\tilde{\mathcal{P}}_{\gamma_\mathcal{P}}(f)(Z),
      \tilde{\mathcal{P}}_{\gamma_\mathcal{P}}(g)(Z)) = |\sup f[Z] -
      \sup g[Z]| \le \sup_{z\in Z}|f(z)-g(z)| \\
      & \le & \sup_{x\in X}
      |f(x)-f(y)| = d_e^\infty(f,g)
    \end{eqnarray*}
    (cf. Lemma~\ref{lem:min-max-ms-de}(\ref{lem:mmmd-2})). The same
    holds with Lemma~\ref{lem:min-max-ms-de}(\ref{lem:mmmd-4}) if we
    replace $d_e$ by $d_\ominus$ and this implies non-expansiveness.
    
    Furthermore we obtain $\omega$-continuity via
    \begin{eqnarray*}
      && \tilde{\mathcal{P}}_{\gamma_\mathcal{P}}(\sup_{i<\omega}
      f_i)(Z) = \sup (\sup_{i<\omega} f_i)[Z] = \sup_{z\in Z}
      \sup_{i<\omega} f_i(z) = \sup_{i<\omega} \sup_{z\in Z} f_i(z) =
      \sup_{i<\omega} f_i[Z] \\
      & = & \tilde{\mathcal{P}}_{\gamma_\mathcal{P}}(f_i)(Z).
    \end{eqnarray*}
    
  \item Observe that
    $\tilde{\mathcal{D}}_{\gamma_\mathcal{D}}(f)\colon
    \mathcal{D}(X)\to [0,1]$ satisfies
    $\tilde{\mathcal{D}}_{\gamma_\mathcal{D}}(f)(p) = \sum_{x\in
      X}f(x)\cdot p(x)$ where $p\in\mathcal{D}X$ is a probability
    distribution on $X$. 
    
    We have
    $d_e(\tilde{\mathcal{D}}_{\gamma_\mathcal{D}}(f)(p),
    \tilde{\mathcal{D}}_{\gamma_\mathcal{D}}(g)(p)) = |\sum_{x\in
      X}f(x)\cdot p(x) - \sum_{x\in X}g(x)\cdot p(x)| \leq \sum_{x\in
      X}p(x)\cdot |f(x) - g(x)| \le 1\cdot d_e^\infty(f,g)$. The same
    holds if we replace $d_e$ by $d_\ominus$ and this implies
    non-expansiveness.
    
    Furthermore we obtain $\omega$-continuity via
    $\tilde{\mathcal{D}}_{\gamma_\mathcal{D}}(\sup_{i<\omega} f_i)(x)
    = \sum_{x\in X}(\sup_{i<\omega} f_i)(x)\cdot p(x) =
    \sup_{i<\omega} \sum_{x\in X}f_i(x)\cdot p(x)$. For the last
    equality we have to exchange the sum and the supremum, for which
    we use the fact that 
    \begin{eqnarray*}
      \sum_{x\in X} f(x) = \sup_{\substack{X'\subseteq X\\X'\text{
            finite}}} \sum_{x\in X'} f(x).
    \end{eqnarray*}
  \item Observe that
    $\tilde{\mathcal{M}}_{\gamma_\mathcal{M}}(f)\colon [0,\top]\to
    [0,\top]$ satisfies
    $\tilde{\mathcal{M}}_{\gamma_\mathcal{M}}(f)(r) = r$. Hence
    $\tilde{\mathcal{M}}_{\gamma_\mathcal{M}}$ maps every function to
    the identity, which is clearly non-expansive and
    $\omega$-continuous.
  \item Observe that
    $\tilde{\mathcal{S}}_{\gamma_\mathcal{S}}(f)\colon \{\bullet\}\to
    [0,\top]$ satisfies
    $\tilde{\mathcal{S}}_{\gamma_\mathcal{S}}(f)(\bullet) =
    \top$. Hence $\tilde{\mathcal{S}}_{\gamma_\mathcal{S}}$ maps every
    function to the constant $\top$-function, which is again clearly
    non-expansive and $\omega$-continuous.
  \end{itemize}
  
\end{proof}

\PropCompEvaluationMaps*

\begin{proof}
  In the following let $f,g\colon X\to[0,\top]$.
  \begin{itemize}
  \item Note that
    $\widetilde{(F\times G)}_\gamma (f) = \tilde{F}_{\gamma_F}(f)
    \circ \pi_1$.
    
    Hence we obtain non-expansiveness via the supremum metric with
    \begin{eqnarray*}
      && d_e^\infty(\widetilde{(F\times
        G)}_\gamma(f),\widetilde{(F\times G)}_\gamma(g)) =
      d_e^\infty(\tilde{F}_{\gamma_F}(f)\circ
      \pi_1, \tilde{F}_{\gamma_F}(g)\circ\pi_1) \\
      & = & d_e^\infty(\tilde{F}_{\gamma_F}(f),
      \tilde{F}_{\gamma_F}(g)) \le d_e^\infty(f,g).
    \end{eqnarray*}
    The same holds if we replace $d_e$ by $d_\ominus$.
    
    Furthermore $\omega$-continuity can be shown as follows:
    \begin{eqnarray*}
      && \widetilde{(F\times G)}_\gamma(\sup_{i<\omega} f_i) =
      \tilde{F}_{\gamma_F}(\sup_{i<\omega} f_i)\circ \pi_1 =
      \sup_{i<\omega} \tilde{F}_{\gamma_F}(f_i)\circ \pi_1 =
      \sup_{i<\omega} \widetilde{(F\times G)}_\gamma(f_i).
    \end{eqnarray*}
  \item Note that
    $\widetilde{(F + G)}_\gamma(f)(t) = \tilde{F}_{\gamma_F}(f)(t)$ if
    $t\in FX$ and $\widetilde{(F + G)}_\gamma(f)(t) = 0$ if $t\in GX$.
    
    Hence we can deduce non-expansiveness via the supremum metric by
    observing that
    $d_e(\widetilde{(F\times G)}_\gamma(f)(t),\widetilde{(F\times
      G)}_\gamma(g)(t))$ is either
    \begin{eqnarray*}
      d_e(\tilde{F}_{\gamma_F}(f)(t),\tilde{F}_{\gamma_F}(g)(t))
      \le d_e^\infty(f,g),
    \end{eqnarray*}
    if $t\in FX$, or the value equals~$0$, if $t\in GX$. The same
    holds if we replace $d_e$ by $d_\ominus$.
    
    Furthermore $\omega$-continuity can be shown as follows: whenever
    $t\in FX$ we have
    \begin{eqnarray*}
      \widetilde{(F + G)}_\gamma(\sup_{i<\omega} f_i))(t) =
      \tilde{F}_{\gamma_F}(\sup_{i<\omega} f_i)(t) = (\sup_{i<\omega}
      \tilde{F}_{\gamma_F}(f_i))(t) = (\sup_{i<\omega} \widetilde{(F +
        G)}_\gamma(f_i))(t).
    \end{eqnarray*}
    And if $t\in GX$ both values are equal to~$0$.
  \item Note that we have
    $\widetilde{FG}_\gamma(f) = \gamma\circ FGf = \gamma_F \circ
    F\gamma_G \circ FGf = \gamma_F\circ F(\gamma_G\circ Gf) =
    \tilde{F}_{\gamma_F}(\tilde{G}_{\gamma_G}(f))$. 
    
    Hence we obtain non-expansiveness via the supremum metric via
    \begin{eqnarray*}
      && d_e^\infty(\widetilde{FG}_\gamma(f),\widetilde{FG}_\gamma(g))
      = d_e^\infty(\tilde{F}_{\gamma_F}(\tilde{G}_{\gamma_G}(f)),
      \tilde{F}_{\gamma_F}(\tilde{G}_{\gamma_G}(g))) \le
      d_e^\infty(\tilde{G}_{\gamma_G}(f), \tilde{G}_{\gamma_G}(g)) \\
      & \le & d_e^\infty(f,g).
    \end{eqnarray*}
    The same holds if we replace $d_e$ by $d_\ominus$.
    
    Furthermore it is easy to show $\omega$-continuity:
    \begin{eqnarray*}
      && \widetilde{FG}_\gamma(\sup_{i<\omega} f_i) =
      \tilde{F}_{\gamma_F}(\tilde{G}_{\gamma_G}(\sup_{i<\omega} f_i))
      = \tilde{F}_{\gamma_F}(\sup_{i<\omega}
      \tilde{G}_{\gamma_G}(f_i)) =
      \sup_{i<\omega}\tilde{F}_{\gamma_F}( \tilde{G}_{\gamma_G}(f_i)) \\
      & = & \sup_{i<\omega} \widetilde{FG}_\gamma(f_i).
    \end{eqnarray*}
    
  \end{itemize}
  
\end{proof}

\begin{lem}
  \label{lem:nonexp-properties-lifting}
  Let $\Gamma$ be a set of evaluation maps.  Every evaluation map
  $\gamma\in \Gamma$ is non-expansive for the Kantorovich lifting,
  i.e.
  $\gamma\colon (F[0,\top],d_e^{\uparrow \Gamma})\neto
  ([0,\top],d_e)$. 
\end{lem}

\begin{proof}
  We have
  $d_e^{\uparrow \Gamma} (t_1,t_2) = \sup\{d_e(\tilde{F}_\gamma
  f(t_1),\tilde{F}_\gamma f(t_2)) \mid f\colon (X,d) \neto ([0,\top],
  d_e),  \gamma\in\Gamma \}$ and we substitute the identity function $\mathit{id}$ for
  $f$. Thus we obtain the following inequality
  $d_e^{\uparrow \Gamma} (t_1,t_2) \geq d_e(\tilde{F}_\gamma
  \mathit{id}(t_1),\tilde{F}_\gamma \mathit{id}(t_2)) =
  d_e(\gamma(t_1),\gamma(t_2)) $, because
  $\tilde{F}_\gamma \mathit{id} = \gamma \circ Fid = \gamma$.
\end{proof}

\PropLiftingPlusDelta*

\begin{proof}
  We set $\delta = d_e^\infty(d_1,d_2)$.  Let $f$ be a function which
  is non-expansive for $d_1$ and $d_e$, i.e.,
  $f\colon (X,d_1)\neto ([0,\top],d_e)$. We define another function
  $h\colon X\to [0,\top]$ via
  $h(z) = \sup\{f(u) - d_2(u,z)\mid u\in X\}$ as in
  Lemma~\ref{lem:approx-nonexpansive}. We know that $f\le h$ and
  $f\colon (X,d_2)\neto([0,\top],d_e)$. Now define
  $g = h - \frac{\delta}{2}$. We have for every $z\in X$:
  \begin{eqnarray*}
    f(z) - g(z) & = & f(z) - h(z) + \frac{\delta}{2} \le
    \frac{\delta}{2} \\
    g(z) - f(z) & = & h(z) - \frac{\delta}{2} - f(z) = \sup\{f(u) -
    d_2(u,z)\mid u\in X\} - \frac{\delta}{2} - f(z) \\
    & = & \sup\{f(u) - f(z) - d_2(u,z)\mid u\in X\} - \frac{\delta}{2} \\
    & \le & \sup\{d_1(u,z) - d_2(u,z)\mid u\in X\} - \frac{\delta}{2}
    \\
    & \le & \delta - \frac{\delta}{2} = \frac{\delta}{2}
  \end{eqnarray*}
  Hence $d_e^\infty(f,g) \le \frac{\delta}{2}$.  Non-expansiveness of
  the predicate lifting wrt.\ the supremum metric
  (cf. Definition~\ref{def:properties-ev-fct}) implies
  $d_e(\tilde{F}_\gamma f(t),\tilde{F}_\gamma h(t)) \le
  \frac{\delta}{2}$ for all $t\in FX$.
  
  Given $t_1,t_2\in FX$ we can infer with the triangle inequality:
  \begin{eqnarray*}
    d_e(\tilde{F}_\gamma f(t_1),\tilde{F}_\gamma f(t_2)) & \le &
    d_e(\tilde{F}_\gamma f(t_1),\tilde{F}_\gamma h(t_1)) +
    d_e(\tilde{F}_\gamma h(t_1),\tilde{F}_\gamma h(t_2)) +
    d_e(\tilde{F}_\gamma h(t_2),\tilde{F}_\gamma f(t_2)) \\
    & \le & d_e(\tilde{F}_\gamma h(t_1),\tilde{F}_\gamma h(t_2)) +
    \delta
  \end{eqnarray*}
  
  Finally:
  \begin{eqnarray*}
    && d_1^{\uparrow \Gamma}(t_1,t_2) \\
    & = & \sup \{d_e(\tilde{F}_\gamma f(t_1),\tilde{F}_\gamma f(t_2))
    \mid f\colon (X,d_1)\neto ([0,\top],d_e),
    \gamma\in\Gamma \} \\
    & = & \sup \{d_e(\tilde{F}_\gamma f(t_1),\tilde{F}_\gamma f(t_2))
    \mid f\colon (X,d_1)\neto ([0,\top],d_e),
    \gamma\in\Gamma, \\
    && \qquad d_e^\infty(f,g)\le \frac{\delta}{2} \text{ for some } g
    \colon
    (X,d_2)\neto ([0,\top],d_e) \} \\
    & \le & \sup \{d_e(\tilde{F}_\gamma g(t_1),\tilde{F}_\gamma
    g(t_2)) + \delta \mid g\colon (X,d_2)\neto ([0,\top],d_e),
    \gamma\in\Gamma \} \\
    & = & \sup \{d_e(\tilde{F}_\gamma g(t_1),\tilde{F}_\gamma g(t_2))
    \mid g\colon (X,d_2)\neto ([0,\top],d_e), \gamma\in\Gamma \} +
    \delta \\
    & = & d_2^{\uparrow \Gamma}(t_1,t_2) + \delta
  \end{eqnarray*}
  Analogously we can show that
  $d_2^{\uparrow \Gamma}(t_1,t_2) \le d_1^{\uparrow \Gamma}(t_1,t_2) +
  \delta$ and this implies $d_e^\infty(d_1^{\uparrow
    \Gamma},d_2^{\uparrow \Gamma})\le \delta$.
  
  In the contractive case the proof is analogous.
\end{proof}

\PropFixpointOmegaSteps*

\begin{proof}~
  
  First note that $d_\omega$ as the pointwise supremum of
  pseudometrics, is again a pseudometric.
  \begin{itemize}
  \item We assume that every $\tilde{F}_\gamma$ is $\omega$-continuous
    and $F$ is $\omega$- accessible. Given $x,y\in FX$, we have to
    prove that:
    \begin{eqnarray*}
      d_\omega^{\uparrow \Gamma}(\alpha(x),\alpha(y)) & = &
      \sup\{d_e(\tilde{F}_\gamma f(\alpha(x)),\tilde{F}_\gamma
      f(\alpha(y))) \mid f\colon (X,d_\omega)\neto
      ([0,\top],d_e),\gamma\in \Gamma \} \\
      & \le & d_\omega(x,y)
    \end{eqnarray*}
    So we have to show that
    $d_e(\tilde{F}_\gamma f(\alpha(x)),\tilde{F}_\gamma f(\alpha(y))
    \le d_\omega(x,y)$ for every non-expansive function
    $f\colon (X,d_\omega)\neto ([0,\top],d_e)$.
    
    Since $F$ is $\omega$-accessible, there is a finite set
    $Z\subseteq X$ with $\alpha(x),\alpha(y)\in FZ$. We can assume
    that $Z$ is non-empty. Now define $g\colon X\to [0,\top]$ as
    $g(z) = f(z)$ if $z\in Z$ and $g(z) = \top$ otherwise. Note that
    $Ff$ and $Fg$ agree on $\alpha(x),\alpha(y)$.
    
    We approximate the function $g$ on $Z$ with an ascending chain of
    functions $g_0\le g_1\le g_2 \le \dots$ with
    $g_i\colon (Z,d_i) \neto ([0,\top],d_e)$, i.e., each $g_i$ is
    non-expansive for $d_i$.  We define
    $g_i(z) = \inf\{g(u)+d_i(u,z)\mid u\in X\}$ as in
    Lemma~\ref{lem:approx-nonexpansive}, which means that $g_i\le g$
    and every $g_i$ is non-expansive as desired. Since
    $d_i\le d_{i+1}$ it also follows that $g_i\le g_{i+1}$.
    
    Since $g(u)=\top$ if $u\not\in Z$, we know that for $z\in Z$ the
    infimum is reached for $u\in Z$ and hence
    $g_i(z) = \inf\{g(u)+d_i(u,z)\mid u\in Z\}$.
    
    Furthermore whenever $z\in Z$
    \begin{eqnarray*}
      g(z) - g_i(z) & = & g(z) - \inf\{g(u)+d_i(u,z)\mid u\in Z\}
      \\
      & = & \sup\{g(z) - g(u) - d_i(u,z)\mid u\in Z\}
      \\
      & = & \sup\{f(z) - f(u) - d_i(u,z)\mid u\in Z\}
      \\
      & \le & \sup\{d_\omega(u,z) - d_i(u,z)\mid u\in Z\}
    \end{eqnarray*}
    Since $Z$ is finite this value converges to $0$ when $i$
    approaches $\omega$ and hence $\sup_{i<\omega} g_i(z) = g(z)$
    whenever $z\in Z$. This means that $\sup_{i<\omega} g_i$,
    $g$ agree on $\alpha(x),\alpha(y)\in FZ$.
    
    Hence, using the fact that $\tilde{F}_\gamma$ is
    $\omega$-continuous and
    Lemma~\ref{lem:min-max-ms-de}(\ref{lem:mmmd-2}):
    \begin{eqnarray*}
      d_e(\tilde{F}_\gamma f(\alpha(x)), \tilde{F}_\gamma f(\alpha(y))
      & = & d_e(\tilde{F}_\gamma g(\alpha(x)),
      \tilde{F}_\gamma g(\alpha(y))) \\
      & = & d_e(\tilde{F}_\gamma (\sup_{i<\omega} g_i)(\alpha(x)),
      \tilde{F}_\gamma (\sup_{i<\omega} g_i)(\alpha(y))) \\
      & = & d_e(\sup_{i<\omega} \tilde{F}_\gamma g_i(\alpha(x)),
      \sup_{i<\omega} \tilde{F}_\gamma g_i(\alpha(y))) \\
      & \le & \sup_{i<\omega} d_e(\tilde{F}_\gamma g_i(\alpha(x)),
      \tilde{F}_\gamma g_i(\alpha(y))) \\
      & \le & \sup_{i<\omega} d_i^{\uparrow \Gamma}(\alpha(x),
      \alpha(y)) \\
      & = & \sup_{i<\omega} d_{i+1}(\alpha(x),\alpha(y)) \\
      & = & d_\omega(x,y)
    \end{eqnarray*}
    The inequality is due to the fact that non-expansiveness of $g_i$
    for $d_i$ implies non-expansiveness of $\tilde{F}_\gamma(g_i)$ for
    $d_i^{\uparrow\Gamma}$.
  \item We assume that every $\tilde{F}_\gamma$ is contractive wrt.\
    the supremum metric (for some constant $c$). Due to
    Lemma~\ref{prop:lifting-nonexpansive} we have that for two metrics
    $d_1,d_2\colon X\times X\to [0,\top]$ it holds that
    \[ d_e^\infty(d_1^{\uparrow \Gamma},d_2^{\uparrow \Gamma}) \le
      c\cdot d_e^\infty(d_1,d_2) \] where $0 < c < 1$.
    
    Note that $\top \neq \infty$, which means that the set of all such
    real-valued bounded functions forms a complete metric space wrt.\
    the supremum metric. 
    Hence we obtain the fixpoint in $\omega$ steps via the Banach
    fixpoint theorem.
  \end{itemize}
  
\end{proof}

\subsubsection{Modal Logics  for the Metric Case}

\TheoHmRVLogicSound*

\begin{proof}~
  \begin{itemize}
  \item We prove (\ref{prop:hrls-1}) and~(\ref{prop:hrls-2}) jointly
    by induction on $i$.
    \begin{itemize}
    \item $i=0$: It is easy to see that for every $\phi$ with
      $md(\phi)=0$ the function $\llbracket \phi \rrbracket$ is
      constant. Due to this
      $d_0^L(x,y) = \sup\{d_e(\llbracket \phi \rrbracket(x),
      \llbracket \phi \rrbracket(y)) \mid md(\phi)=0\}=0 \le
      d_0(x,y)$. All constant functions are non-expansive for $d_0$.
    \item $i\to i+1$: We first show that for every $\phi$ with
      $\mathit{md}(\phi) \le i+1$ that
      $\llbracket\phi\rrbracket\colon (X,d_{i+1})\neto ([0,\top],d_e)$
      is non-expansive by structural induction on $\phi$.
      \begin{itemize}
      \item $\phi=\top$: This case can not occur for
        $\mathit{md}(\phi) > 0$.
      \item $\phi= [\gamma]\psi$: Here $\mathit{md}(\psi)\le i$ and
        $\llbracket [\gamma]\psi \rrbracket = \gamma \circ F\llbracket
        \psi \rrbracket \circ \alpha$, where we have a composition of
        non-expansive functions:
        \begin{itemize}
        \item The evaluation map
          $\gamma\colon (F[0,\top],d_e^{\uparrow \Gamma})\neto
          ([0,\top],d_e)$ is non-expansive by
          Lemma~\ref{lem:nonexp-properties-lifting}.
        \item
          $\llbracket \psi \rrbracket\colon (X,d_i)\neto ([0,1],d_e)$
          is non-expansive by the outer induction hypothesis and,
          since the lifting preserves non-expansive functions,
          $F\llbracket \psi \rrbracket\colon (FX,d_i^{\uparrow
            \Gamma})\neto (F[0,1],d_e^{\uparrow \Gamma})$.
        \item By definition
          $d_{i+1} = d_i^{\uparrow \Gamma} \circ (\alpha\times \alpha)$,
          hence
          $\alpha\colon (X,d_{i+1}) \neto (FX,d_i^{\uparrow
            F})$ (it is even an isometry).
        \end{itemize}
      \item $\phi= \min(\psi,\psi')$: We know that
        $\mathit{md}(\psi),\mathit{md}(\psi')\le \mathit{md}(\phi)\le
        i+1$. The inner induction hypothesis implies that
        $\llbracket \psi\rrbracket$, $\llbracket\psi'\rrbracket$ are
        both non-expansive for $d_{i+1}$. Using
        Lemma~\ref{lem:min-max-ms-de}(\ref{lem:mmmd-3}) we know that
        for all $x,y \in X$
        \begin{eqnarray*}
          d_e(\llbracket\phi\rrbracket(x),\llbracket\phi\rrbracket(y))
          & = &
          d_e(\min\{\llbracket\psi\rrbracket(x),\llbracket\psi'\rrbracket(x)\},
          \min\{\llbracket\psi\rrbracket(y),\llbracket\psi'\rrbracket(y)\})
          \\
          & \leq &
          \max\{d_e(\llbracket\psi\rrbracket(x),\llbracket\psi\rrbracket(y)),
          d_e(\llbracket\psi'\rrbracket(x),\llbracket\psi'\rrbracket(y))\}
          \leq d_{i+1}(x,y)
        \end{eqnarray*}
      \item $\phi(x)=\lnot\psi(x)$: We know that
        $\mathit{md}(\psi) = \mathit{md}(\phi)\le i+1$. The inner
        induction hypothesis implies that $\llbracket \psi\rrbracket$
        is non-expansive for $d_{i+1}$. For $x,y \in X$ we obtain
        \begin{eqnarray*}
          d_e(\llbracket \phi \rrbracket(x),\llbracket \phi
          \rrbracket(y)) & = & d_e(\llbracket \top-\psi
          \rrbracket(x),\llbracket \top-\psi
          \rrbracket(y))=|\top-\llbracket\psi
          \rrbracket(x)-(\top-\llbracket\psi\rrbracket(y))| \\
          & = &
          |\llbracket\psi\rrbracket(y)-\llbracket\psi\rrbracket(x)| =
          d_e(\llbracket \psi \rrbracket(x), \llbracket \psi
          \rrbracket(y)) \leq d_{i+1}(x,y)
        \end{eqnarray*}
      \item $\phi(x)=\psi \ominus q$: We know that
        $\mathit{md}(\psi) = \mathit{md}(\phi)\le i+1$. The inner
        induction hypothesis implies that $\llbracket \psi\rrbracket$
        is non-expansive for $d_i$. For all $x,y \in X$ we know by
        Lemma~\ref{lem:min-max-ms-de}(\ref{lem:mmmd-1}) that
        \[
          d_e(\llbracket \phi \rrbracket(x),\llbracket \phi
          \rrbracket(y)) = d_e(\llbracket \psi \rrbracket(x) \ominus q
          ,\llbracket \psi \rrbracket(y) \ominus q )\leq
          d_e(\llbracket \psi \rrbracket(x), \llbracket \psi
          \rrbracket(y)) \leq d_{i+1}(x,y)
        \]
      \end{itemize}
      Thus, we conclude by the non-expansiveness of all formulas
      $\phi$ with $\mathit{md}(\phi)\leq i$ that
      $d_i^L(x,y)= \sup\{ d_e(\llbracket \phi \rrbracket (x)
      ,\llbracket \phi \rrbracket (y)\ | \ \mathit{md}(\phi)\leq
      i \} \leq d_i(x,y)$.
    \end{itemize}
  \item (\ref{prop:hrls-3}) follows directly from (\ref{prop:hrls-1}) by
    taking the supremum on both sides and observing that
    $d_\alpha\ge \sup_{i<\omega} d_i$.
  \end{itemize}
  
\end{proof}

\PropPreservTotalBounded*

\begin{proof}
  We denote the set of all non-expansive functions
  $f\colon (X,d) \xrightarrow{\text{1}} ([0,\top],d_e)$ with
  $\mathcal{F}$. We know from \cite[Lemma 5.6]{wspk:van-benthem-fuzzy}
  that if $(X,d)$ is a totally bounded space, then $\mathcal{F}$ is
  also totally bounded wrt.\ the supremum metric $d_e^\infty$. This is
  a variation of the Arzel\`{a}a-Ascoli theorem.\footnote{Actually this lemma is stated for $\top = 1$, but the proof can be
    straightforwardly adapted.} Due to this for all $\epsilon >0$
  there exists a finite set $\{f_1, \dots, f_n\}$ with
  $f_i \in \mathcal{F}$ such that for all $f \in \mathcal{F}$ we have
  one $f_i$ with $d_e^\infty(f,f_i)\leq \epsilon$. Given a fixed
  $\epsilon$, we denote the function $f_i$ corresponding to $f$ by
  $\hat{f}$.
  
  Furthermore, since the predicate lifting is non-expansive wrt.\ the
  supremum metric (cf. Definition~\ref{def:properties-ev-fct}), we know
  that for all $t \in FX$
  $d_e(\tilde{F}_\gamma f(t),\tilde{F}_\gamma \hat{f}(t)) \leq
  \epsilon$.  And we have, using the triangle inequality,
  \begin{align*}
    d^{\uparrow \Gamma}(t_1,t_2) & = \sup\{d_e(\tilde{F}_\gamma
    f(t_1),\tilde{F}_\gamma f(t_2)) \mid f \in
    \mathcal{F},\gamma\in\Gamma\}\\
    & \leq \sup\{d_e(\tilde{F}_\gamma f(t_1),\tilde{F}_\gamma
    \hat{f}(t_1)) + d_e(\tilde{F}_\gamma \hat{f}(t_1),\tilde{F}_\gamma
    \hat{f}(t_2)) +
    d_e(\tilde{F}_\gamma \hat{f}(t_2),\tilde{F}_\gamma f(t_2)) \mid \\
    & \qquad\qquad f \in
    \mathcal{F},\gamma\in\Gamma \}\\
    &\leq \sup\{d_e(\tilde{F}_\gamma \hat{f}(t_1),\tilde{F}_\gamma
    \hat{f}(t_2)) \mid
    f \in \mathcal{F},\gamma\in\Gamma\} + 2 \cdot \epsilon \\
    &= \sup\{ d_e(\tilde{F}_\gamma f_i(t_1),\tilde{F}_\gamma f_i(t_2))\mid i \in
    \{1,\dots,n\}, \gamma\in\Gamma \} + 2 \cdot \epsilon
  \end{align*} 
  By assumption $\Gamma$ is finite and we assume
  $\Gamma = \{\gamma_1,\dots,\gamma_k\}$.  Now we define
  $g \colon FX \to [0,\top]^{k\cdot n}$ with
  \[ g(t)=(\tilde{F}_{\gamma_1} f_1(t),\dots,\tilde{F}_{\gamma_1}
    f_n(t), \tilde{F}_{\gamma_k} f_1(t),\dots,\tilde{F}_{\gamma_k}
    f_n(t)). \] Since $[0,\top]^{k\cdot n}$ is compact under the
  supremum metric there exists finitely many
  $u_1,\dots,u_m \in [0,\top]^{k\cdot n}$ such that
  $\bigcup_{i=1}^m B_\epsilon(u_i)= [0,\top]^{k\cdot n}$. The preimages of all
  balls cover $FX$ and each preimage $g^{-1}(B_\epsilon(u_i))$ has a
  diameter at most $4\cdot \epsilon$: For
  $s_1,s_2 \in g^{-1}(B_\epsilon(u_i))$ it holds that:
  \begin{align*}
    d^{\uparrow \Gamma}(s_1,s_2) & \leq
    2 \cdot \epsilon + \sup_{i \in \{1,\dots,n\}, \gamma\in\Gamma}
    \{d_e(\tilde{F}_\gamma f_i(s_1),\tilde{F}_\gamma f_i(s_2))\}\\
    & = 2 \cdot \epsilon + d_e^\infty(g(s_1),g(s_2))\\
    & \leq 2 \cdot \epsilon +  d_e^\infty(g(s_1),u_i) +  
    d_e^\infty(u_i,g(s_2))\\
    & \leq 2 \cdot \epsilon + 2 \cdot \epsilon = 4\cdot  \epsilon
  \end{align*} 
  Hence, given a $\delta > 0$ we can set
  $\epsilon = \frac{\delta}{4}$ and obtain balls
  $g^{-1}(B_\epsilon(u_i))$ of diameter at most $\delta$, which cover
  $FX$.
\end{proof}

\PropApproxMetricsTotallyBounded*

\begin{proof}
  We proceed by induction on $i$ and start with the trivial (constant)
  pseudometric $d_0$, hence$(X,d_0)$ is a totally bounded pseudometric
  space, since $d_0(x,y) =0 < \epsilon $ for all $\epsilon >0$.  
  
  For $(X,d_{i+1})$ we know from the induction hypothesis and by
  Proposition~\ref{prop:preserv-total-bounded} that
  $(FX,d_i ^{\uparrow \Gamma})$ is totally bounded. So, for every
  $\epsilon>0$, there exist $t_1,\cdots,t_n \in FX$ such that
  $FX = \bigcup^n_{i=1} B_\frac{\epsilon}{2}(t_i)$, which implies
  $X = \bigcup^n_{i=1}\alpha^{-1}(B_\frac{\epsilon}{2}(t_i))$. Let
  $x,y \in \alpha^{-1}(B_\frac{\epsilon}{2}(t_i))$. Then $x,y$ have at
  most distance $\epsilon$:
  $d_{i+1}(x,y)= d_i^{\uparrow \Gamma} (\alpha(x),\alpha(y)) \le
  d_i^{\uparrow \Gamma} (\alpha(x),t_i) + d_i^{\uparrow \Gamma}
  (t_i,\alpha(y)) \leq \epsilon$.
\end{proof}

\PropFormulasDenseNonexpansive*

\begin{proof}
  We have to show that for every $\epsilon > 0$ and every
  non-expansive function $f \colon (X,d_i^L) \neto ([0,\top],d_e)$,
  there exists formula $\phi$ with $\mathit{md}(\phi)\le i$ and
  $d_e^\infty(f,\llbracket \phi\rrbracket) \le \epsilon$.
  
  From Proposition~\ref{prop:approx-metrics-totally-bounded} we know
  that $(X,d_i)$ is totally bounded and since $d_i^L\le d_i$
  (Theorem~\ref{prop:hm-rv-logic-sound}) we can infer that $(X,d_i^L)$
  is also totally bounded (since
  $B_\epsilon^{d_i}(x)\subseteq B_\epsilon^{d_i^L}(x)$).
  
  We use the following result from \cite[Lemma
  5.8]{wspk:van-benthem-fuzzy}, which is a variation of a lemma by Ash
  \cite[Lemma A.7.2]{a:real-analysis-probability}, adapted from
  compact spaces and continuous functions to totally bounded spaces
  and non-expansive functions.\footnote{Actually this lemma is
    stated for $\top = 1$, but the proof can be straightforwardly
    adapted.}
  
  \begin{quote}
    Let $(X,d)$ be a totally bounded pseudometric space and let
    $\mathcal{G}$ be a subset of
    $\mathcal{F} = \{f\colon (X,d)\neto ([0,\top],d_e)$ such that
    $f_1,f_2\in \mathcal{G}$ implies
    $\min\{f_1,f_2\},\max\{f_1,f_2\}\in \mathcal{G}$. If each
    $f\in \mathcal{F}$ can be approximated at each pair of points by
    functions in $\mathcal{G}$, then $\mathcal{G}$ is dense in
    $\mathcal{F}$ (wrt.\ $d_e^\infty$).
  \end{quote}
  
  Here $d = d_i^L$ and $\mathcal{G}$ is the set of functions
  $\llbracket\phi\rrbracket$, where $\mathit{md}(\phi)\le i$. Since
  $\min$ and $\lnot$ are operators of the logic and neither increases
  the modal depth, $\mathcal{G}$ is
  clearly closed under $\max$ and $\min$.
  
  Now let $f\colon (X,d)\neto ([0,\top],d_e)$, $\delta > 0$ and
  $x,y\in X$. We have to show that there exists a modal formula $\phi$
  (with modal depth at most $i$) that approximates $f$ at these
  points. That is $d_e(f(x),\llbracket\phi\rrbracket(x)) \le \delta$
  and $d_e(f(y),\llbracket\phi\rrbracket(y)) \le \delta$.
  
  We concentrate on the case $f(x) \geq f(y)$, the other case is
  analogous.
  \[
    \varDelta := f(x)-f(y) \leq d_i^L(x,y) = \sup \{d_e(\llbracket
    \phi \rrbracket(x),\llbracket \phi \rrbracket(y)) \mid
    \mathit{md}( \phi) \leq i \} \] Then there exists a formula $\psi$
  with $\mathit{md}(\psi)\leq i$ such that
  $\varDelta -\delta \leq \llbracket \psi \rrbracket(x) - \llbracket
  \psi \rrbracket(y)$ (whenever we find $\psi$ with
  $\varDelta -\delta \leq \llbracket \psi \rrbracket(y) - \llbracket
  \psi \rrbracket(x)$ we use negation). We can assume that
  $\llbracket\psi\rrbracket(y) \le \llbracket\psi\rrbracket(x)$.
  
  Whenever $\llbracket\psi\rrbracket(y) \ge f(y)$ we set
  $ \phi = \min(\psi\ominus r,s)$ with
  $r,s \in\mathbb{Q}\cap [0,\top]$ chosen as follows:
  \begin{eqnarray*}
    r & \in & [\llbracket \psi \rrbracket(y) - f(y) - \delta,
    \llbracket \psi \rrbracket(y) - f(y)] \\
    s & \in & [f(x),f(x)+\delta]
  \end{eqnarray*}
  Since the depth of a formula only increases with the modality
  operator, it holds that $\mathit{md}(\phi)\le i$.
  
  We show that $\phi$ is the required formula. First note that
  $r\le \llbracket\psi\rrbracket(y) \le \llbracket\psi\rrbracket(x)$
  and $f(y)\le f(x)\le s$.
  \begin{eqnarray*}
    \llbracket\phi\rrbracket(y) - f(y) & = &
    \min\{\llbracket\psi\rrbracket(y)-r,s\}-f(y) =
    \min\{\underbrace{\llbracket\psi\rrbracket(y)-f(y)-r}_{\le
      \delta},s-f(y)\} \le \delta \\
    f(y) - \llbracket\phi\rrbracket(y) & = & f(y) -
    \min\{\llbracket\psi\rrbracket(y)-r,s\} =
    \max\{\underbrace{f(y)-\llbracket\psi\rrbracket(y)+r}_{\le
      0},\underbrace{f(y)-s}_{\le 0}\} \le \delta \\
    \llbracket\phi\rrbracket(x) - f(x) & = &
    \min\{\llbracket\psi\rrbracket(x)-r,s\}-f(x) =
    \min\{\llbracket\psi\rrbracket(x)-f(x)-r,\underbrace{s-f(x)}_{\le
      \delta}\} \le \delta \\
    f(x) - \llbracket\phi\rrbracket(x) & = & f(x) -
    \min\{\llbracket\psi\rrbracket(x)-r,s\} =
    \max\{\underbrace{f(x)-\llbracket\psi\rrbracket(x)+r}_{\le
      \delta},\underbrace{f(x)-s}_{\le 0}\} \le \delta 
  \end{eqnarray*}
  $f(x)-\llbracket\psi\rrbracket(x)+r\le \delta$ holds since
  $\llbracket\psi\rrbracket(x) - \llbracket\psi\rrbracket(y)\ge
  \Delta-\delta = f(x)-f(y)-\delta$. Then
  $f(x)-\llbracket\psi\rrbracket(x)+r \le
  f(y)-\llbracket\psi\rrbracket(y) + \delta +
  \llbracket\psi\rrbracket(y) - f(y) = \delta$.
  
  Whenever $\llbracket\psi\rrbracket(y) \le f(y)$ we define
  $\phi = \min(\psi\oplus r,s)$ where
  $r\in [f(y) - \llbracket \psi \rrbracket(y) - \delta, f(y) -
  \llbracket \psi \rrbracket(y)]$ and $s$ is chosen as above. Note
  that $\oplus$ with $a\oplus b = \min\{a+b,\top\}$ is not an operator
  of the logic, but it can be simulated by
  $\lnot (\lnot a \ominus b)$.
\end{proof}

\ThmHmRVLogicExpressive*

\begin{proof}
  We will show that $d^L$ is a prefixpoint of the equation
  $d = d^{\uparrow \Gamma} \circ (\alpha \times \alpha)$.  Since the
  fixpoint $d_\alpha$ is reached in $\omega$ steps (i.e.,
  $d_\alpha = \sup_{i<\omega} d_i$) it suffices to show
  $d_i \leq d_i^L$ by induction on $i$:
  \begin{itemize}
  \item $i=0$: The inequality is clearly satisfied since $d_0 = 0$.
  \item $i \to i+1$: Let $x,y\in X$.  We first observe that for every
    $\epsilon > 0$ and $f\colon (X,d_i^L) \neto ([0,\top],de)$ there
    exists a modal formula $\psi$ with $\mathit{md}(\psi) \le i$ and
    $d_e^\infty(f,\llbracket \psi\rrbracket) \le \epsilon$
    (cf. Proposition~\ref{prop:formulas-dense-nonexpansive}). With the
    triangle inequality we have
    \begin{eqnarray*}
      && d_e(\tilde{F}_\gamma f(\alpha(x)),\tilde{F}_\gamma
      f(\alpha(y))) \\
      & \le & d_e(\tilde{F}_\gamma f(\alpha(x)),\tilde{F}_\gamma
      \llbracket\psi\rrbracket(\alpha(x))) + d_e(\tilde{F}_\gamma
      \llbracket\psi\rrbracket(\alpha(x)),\tilde{F}_\gamma
      \llbracket\psi\rrbracket(\alpha(y))) \\
      && \qquad + d_e(\tilde{F}_\gamma
      \llbracket\psi\rrbracket(\alpha(y)),\tilde{F}_\gamma
      f(\alpha(y))) 
    \end{eqnarray*}
    Non-expansiveness of the predicate lifting wrt.\ to the supremum
    metric (cf. Definition~\ref{def:properties-ev-fct}) implies that
    $d_e(\tilde{F}_\gamma f(\alpha(x)),\tilde{F}_\gamma
    \llbracket\psi\rrbracket(\alpha(x))) \le \epsilon$, analogously
    $d_e(\tilde{F}_\gamma
    \llbracket\psi\rrbracket(\alpha(y)),\tilde{F}_\gamma f(\alpha(y)))
    \le \epsilon$. Combined we obtain
    \[ d_e(\tilde{F}_\gamma f(\alpha(x)),\tilde{F}_\gamma
      f(\alpha(y))) \le d_e(\tilde{F}_\gamma
      \llbracket\psi\rrbracket(\alpha(x)),\tilde{F}_\gamma
      \llbracket\psi\rrbracket(\alpha(y))) + 2\cdot \epsilon. \]
    This means that 
    \begin{eqnarray*}
      && \sup\{d_e(\tilde{F}_\gamma f(\alpha(x)),\tilde{F}_\gamma
      f(\alpha(y))) \mid \ f\colon (X,d_i^L) \neto ([0,\top],de),
      \gamma\in\Gamma\} \\ 
      & \le & \sup\{d_e(\tilde{F}_\gamma\llbracket \psi
      \rrbracket(\alpha(x)), \tilde{F}_\gamma\llbracket \psi
      \rrbracket(\alpha(y))) \mid \mathit{md}(\psi) \leq i,
      \gamma\in\Gamma \} + 2\cdot \epsilon 
    \end{eqnarray*}
    and since this holds for every $\epsilon > 0$, the two suprema are
    equal.
    \begin{eqnarray*}
      d_{i+1}(x,y) &=& \sup\{d_e(\tilde{F}_\gamma
      f(\alpha(x)),\tilde{F}_\gamma f(\alpha(y))) \mid f\colon (X,d_i)
      \neto ([0,\top],de), \gamma\in\Gamma\}\\
      & &(\text{by the induction hypothesis } d_i \leq d_i^L)\\
      & \leq & \sup\{d_e(\tilde{F}_\gamma
      f(\alpha(x)),\tilde{F}_\gamma f(\alpha(y))) \mid \ f\colon
      (X,d_i^L) \neto
      ([0,\top],de), \gamma\in\Gamma\}\\
      & = & \sup\{d_e(\tilde{F}_\gamma\llbracket \psi
      \rrbracket(\alpha(x)),
      \tilde{F}_\gamma\llbracket \psi \rrbracket(\alpha(y))) 
      \mid \mathit{md}(\psi) \leq i, \gamma\in\Gamma \}\\
      & = & \sup\{d_e(\llbracket [\gamma] \psi \rrbracket(x),
      \llbracket [\gamma] \psi \rrbracket(y)) \mid 
      \mathit{md}(\psi) \leq i, \gamma\in\Gamma \}\\
      & \leq & \sup\{d_e(\llbracket \phi \rrbracket(x),\llbracket
      \phi \rrbracket(y))) \mid \mathit{md}(\phi) \leq
      i+1 \}\\
      & = & d_{i+1}^L(x,y)
    \end{eqnarray*}
  \end{itemize}
\end{proof}

\subsubsection{Games for the Metric Case}

\PropGameDistancePseudometric*

\begin{proof} 
  Here we need to review the three conditions of a pseudometric.
  Assume that $x,y,z\in X$.
  
  \begin{enumerate}
  \item \emph{Reflexivity:} we show that there is a strategy for~$D$
    for $(x,x,0)$ and hence $d_\alpha^G(x,x)=0$.
    
    Assume that S chooses $x$ and some $p_1$, then D answers with
    $p_2=p_1$ and also chooses $x$. In this case
    $d_\ominus(\tilde{F}_\gamma p_1(\alpha(x)), \tilde{F}_\gamma
    p_2(\alpha(x))) = 0$ for every $\gamma\in\Gamma$, due to
    reflexivity of $d_\ominus$. S then chooses some $x'$ and D can
    choose the same $x'$, since $p_2(x') = p_1(x')$. Furthermore
    $\epsilon' = p_2(x')-p_1(x') = 0$ and the game continues with
    $(x',x',0)$, i.e., in a situation analogous to the previous one.
  \item \emph{Symmetry:} we show that
    $d_\alpha^G(x,y)=d_\alpha^G(y,x)$.
    
    This is true since S can choose either $x$ or $y$ and
    D is then forced to play with the other state. Hence the
    roles of $x$ and $y$ can be exchanged. Hence
    \begin{eqnarray*}
      d_\alpha^G(x,y) & = & \inf \{\epsilon \mid \text{D has a
        winning strategy
        for } (x,y,\epsilon) \} \\
      & = & \inf \{\epsilon \mid \text{D has a winning
        strategy for } (y,x,\epsilon) \} = d_\alpha^G(y,x)
    \end{eqnarray*}
  \item \emph{Triangle inequality:} We prove that
    $d_\alpha^G(x,z) \leq d_\alpha^G(x,y) + d_\alpha^G(y,z)$. We do
    this by constructing a winning strategy for $D$ for
    $(x,z,\theta+\delta)$ from strategies for
    $(x,y,\theta),(y,z,\delta)$.
    
    Now, assume that in the $(x,z,\theta+\delta)$-game S chooses
    $p_i$, $i\in\{1,3\}$ and $x'$ in Step~1. We can assume that $i=1$
    (the other case is analogous).
    
    If $S$ chooses $p_1,x$ in the $(x,y,\theta)$-game, D can play
    according to her strategy and choose $p_2$ and $y$ with
    $d_\ominus(\tilde{F}_\gamma p_1(\alpha(x)),\tilde{F}_\gamma
    p_2(\alpha(y)))\le \theta$ for all $\gamma$. Now assume that in
    the $(y,z,\delta)$-game S chooses $y$ and $p_2$. Then D has an
    answering move $p_3$ and $z$ with
    $d_\ominus(\tilde{F}_\gamma p_2(\alpha(y)),\tilde{F}_\gamma
    p_3(\alpha(z)))\le \delta$. Combined we have, due to the triangle
    inequality:
    \begin{eqnarray*}
      && d_\ominus(\tilde{F}_\gamma p_1(\alpha(x)),\tilde{F}_\gamma p_3( \alpha(z))) \\
      & \le & d_\ominus(\tilde{F}_\gamma
      p_1(\alpha(x)),\tilde{F}_\gamma p_2(\alpha(y))) +
      d_\ominus(\tilde{F}_\gamma p_2(\alpha(y)),\tilde{F}_\gamma
      p_3(\alpha(z))) \le \theta + \delta
    \end{eqnarray*}
    Hence $D$ will answer S's move in the $(x,z,\theta+\delta)$-game
    by $p_3,z$ in Step~2.
    
    Now assume that S chooses $p_i$ with $i\in\{1,3\}$ and some state
    $x'\in X$ (again we assume without loss of generality that $i=1$)
    in Step~3.  Then, following the strategy for $(x,y,\theta)$, D
    chooses $y'$ with $p_1(x') \le p_2(y')$. Assume that S would
    choose $p_2$ and $y'$ in the $(y,z,\delta)$-game. Following her
    strategy D would choose $z'$ with $p_2(y')\le p_3(z')$. Combined,
    we have $p_1(x') \le p_3(z')$.
    
    Hence $D$ will answer S's move by $z'$ in Step~4.
    
    Furthermore
    $p_3(z')-p_1(x') = \underbrace{p_3(z') - p_2(y')}_{= \theta'} +
    \underbrace{p_2(y') - p_1(x')}_{\delta'}$. Hence we are in the
    situation $(x',z',\theta'+\delta')$ with existing winning
    strategies for the $(x',y',\theta')$- and $(y',z',\delta')$-games.
    
    Finally, we obtain
    \begin{eqnarray*}
      \epsilon_{x,z} & = & \inf \{\epsilon \mid \text{D has a
        winning strategy for } (x,z,\epsilon)\} \\
      & \le & \inf \{\theta+\delta \mid \text{D has a
        winning strategy for } (x,y,\theta) \text{ and } (y,z,\delta)
      \} \\
      & = & d_\alpha^G(x,y) + d_\alpha^G(y,z)
    \end{eqnarray*}
  \end{enumerate}
  
\end{proof}

\ThmWinningStrategyDefenderMetric*

\begin{proof}
  Let $x,y\in X$. We show for all $\delta > 0$ that whenever
  $d_\alpha(x,y)\le \epsilon - \delta$, then D can win the
  $(x,y,\epsilon)$-game. This implies in particular that D has a
  winning strategy for $(x,y,\epsilon)$ with
  $\epsilon = d_\alpha(x,y) + \delta$. Since $d_\alpha^G(x,y)$ is the
  infimum of all such $\epsilon$ we have
  $d_\alpha^G(x,y)\le d_\alpha(x,y) + \delta$ and since this holds for
  all $\delta > 0 $ we get the desired inequality.
  
  Assume that S chooses $s \in \{x,y\}$ with
  $p_1 \colon X \rightarrow [0,\top]$ in Step~1. In this case D chooses
  in Step~2 $p_2 = \max\{p_1,h-\frac{\delta}{2}\}$ where
  $h(z) = \sup \{p_1(u)-d_\alpha(u,z) \mid u \in X\}$.
  Lemma~\ref{lem:approx-nonexpansive} implies that $p_1\le h$ and
  $h\colon (X,d_\alpha)\neto ([0,\top],d_e)$ is non-expansive. Clearly
  $p_1\le p_2$.
  
  The inequality $p_1-p_2\le 0$, i.e.\
  $d_\ominus^\infty(p_1,p_2)\le 0$, implies
  $d_\ominus(\tilde{F}_\gamma p_1(\alpha(s)),\tilde{F}_\gamma
  p_2(\alpha(s))) \le 0$ with the non-expansiveness of the predicate
  lifting (cf. Definition~\ref{def:properties-ev-fct}).
  
  Furthermore we observe that
  $p_2\colon (X,d_\alpha+\frac{\delta}{2})\neto ([0,\top],d_e)$: Let
  $u,z\in X$ and we show that
  $p_2(z) - p_2(u) \le d_\alpha(u,z)+\frac{\delta}{2}$. Whenever
  $p_2(u) = p_1(u)$ we have that $p_1(u) \ge h(u) - \frac{\delta}{2}$
  and hence: 
  \begin{eqnarray*}
    p_2(z) - p_2(u) & = & \max\left\{p_1(z),h(z)-\frac{\delta}{2}\right\} -
    p_1(u) = \max\left\{p_1(z)-p_1(u),h(z)-\frac{\delta}{2}-p_1(u)\right\} \\
    & \le &
    \max\left\{h(z)-h(u)+\frac{\delta}{2},h(z)-\frac{\delta}{2}-h(u)
      +\frac{\delta}{2} \right\} \le d_\alpha(u,z) + \frac{\delta}{2}
  \end{eqnarray*}
  Analogously, whenever $p_2(u) = h(u) - \frac{\delta}{2}$ we have:
  \begin{eqnarray*}
    p_2(z) - p_2(u) & = &
    \max\left\{p_1(z),h(z)-\frac{\delta}{2}\right\} -
    h(u) + \frac{\delta}{2} = \max\left\{p_1(z)-h(u)+\frac{\delta}{2},
      h(z)-h(u)\right\} \\
    & \le & \max\left\{h(z)-h(u)+\frac{\delta}{2},
      h(z)-h(u)\right\} \le d_\alpha(u,z) + \frac{\delta}{2}
  \end{eqnarray*}
  Furthermore we know from $d_\ominus\le d_e$ and the
  non-expansiveness of $\gamma$, $p_2$ (and hence of
  $\tilde{F}_\gamma p_2$) that
  \[
    d_\ominus(\tilde{F}_\gamma p_2(\alpha(s)),\tilde{F}_\gamma
    p_2(\alpha(t))) \le d_e(\tilde{F}_\gamma
    p_2(\alpha(s)),\tilde{F}_\gamma p_2(\alpha(t))) \leq
    \bigg(d_\alpha+\frac{\delta}{2}\bigg)^{\uparrow \Gamma}(\alpha(s),
    \alpha(t)) \] From Lemma~\ref{prop:lifting-nonexpansive} it follows
  that
  \[ \bigg(d_\alpha+\frac{\delta}{2}\bigg)^{\uparrow
      \Gamma}(\alpha(s), \alpha(t)) \le d_\alpha^{\uparrow
      \Gamma}(\alpha(s), \alpha(t)) + \delta = d_\alpha(s,t) + \delta
    \leq \ \epsilon - \delta + \delta = \epsilon \]
  Now the triangle inequality implies
  \[ d_\ominus(\tilde{F}_\gamma p_1(\alpha(s)),\tilde{F}_\gamma
    p_2(\alpha(t))) \le \underbrace{d_\ominus (\tilde{F}_\gamma
      p_1(\alpha(s)), \tilde{F}_\gamma p_2(\alpha(s)))}
    _{\substack{= 0}} + \underbrace{d_\ominus (\tilde{F}_\gamma
      p_2(\alpha(s)),\tilde{F}_\gamma p_2(\alpha(t)))}_{\substack{\le
        \epsilon}} \leq \epsilon, \] hence $p_2$ is a valid choice
  for~D. S now chooses $i$ and $x' \in X$ in Step~3. We distinguish
  the following cases:
  \begin{itemize}
  \item Case $i=1$: D chooses $y'=x'$ in Step~4. We have
    $p_1(x') \leq p_2(x') = p_2(y') $ and
    $ \epsilon' = p_2(x')-p_1(y')\geq 0 = d_\alpha(x',y')$, a
    situation from which D has a winning strategy by copying all moves
    of S.
  \item Case $i=2$: If $p_1(x') = p_2(x')$ D can again choose
    $y' = x'$ in Step~4 and we are in a situation analogous to
    Case~$i=1$ above.
    
    Otherwise $p_2(x') = h(x') - \frac{\delta}{2}$ and there exists a
    $y'\in X$ such that
    $h(x') \le p_1(y') - d_\alpha(x',y') + \frac{\delta}{4}$, which is
    chosen by D in Step~4. It holds that
    $p_2(x') = h(x') - \frac{\delta}{2} \le p_1(y') - d_\alpha(x',y')
    + \frac{\delta}{4} - \frac{\delta}{2} = p_1(y') - d_\alpha(x',y')
    - \frac{\delta}{4} \le p_1(y')$ and
    $\epsilon' = p_1(y')-p_2(x')\ge d_\alpha(x',y') +
    \frac{\delta}{4}$.
    
    Now the game continues with the next round $(x',y',\epsilon')$,
    where $d_\alpha(x',y')\le \epsilon'-\frac{\delta}{4}$ and so D has
    a winning strategy.
    
  \end{itemize}
\end{proof}

\ThmWinningStrategyDefenderMetricRev*

\begin{proof}
  Due to Proposition~\ref{prop:game-distance-pseudometric} we know
  that $d_\alpha^G$ is a pseudometric. We will now show that it is a
  prefix-point of the defining fixpoint equation, i.e.
  $d_\alpha^G \geq (d_\alpha^G)^{\uparrow \Gamma}\circ
  (\alpha\times\alpha)$. Since $d_\alpha$ is the smallest
  (pre-)fixpoint, this implies that $d_\alpha\le d_\alpha^G$.
  
  Let $x,y\in X$.  It holds that
  \[ (d_\alpha^G)^{\uparrow \Gamma}(\alpha(x),\alpha(y)) =
    \sup\{d_e(\tilde{F}_\gamma f(\alpha(x)), \tilde{F}_\gamma
    f(\alpha(y))) \mid f\colon(X,d_\alpha^G)\neto ([0,\top],d_e),
    \gamma\in \Gamma \}. \] We will show that
  $d_\ominus(\tilde{F}_\gamma f(\alpha(x)), \tilde{F}_\gamma
  f(\alpha(y))) \le d_\alpha^G(x,y)$ and the same holds analogously if
  the roles of $x,y$ are reversed. This means that every element in
  the $\sup$ is bounded by $d_\alpha^G(x,y)$, from which the inquality
  follows.  It is sufficient to show that
  $d_\ominus(\tilde{F}_\gamma f(\alpha(x)), \tilde{F}_\gamma
  f(\alpha(y))) \le \epsilon$ whenever D has a winning strategy for
  $(x,y,\epsilon)$.
  
  So, assuming that S chooses $x$ and $f$ at Step~1, we know that D can
  at Step~2 choose a real-valued predicate $p_2$, such that
  $d_\ominus(\tilde{F}_\gamma f(\alpha(x)), \tilde{F}_\gamma
  p_2(\alpha(y))) \leq \epsilon$ for all $\gamma\in\Gamma$.
  
  We infer from the triangle inequality that
  \[ d_\ominus(\tilde{F}_\gamma f(\alpha(x)), \tilde{F}_\gamma
    f(\alpha(y))) \leq d_\ominus(\tilde{F}_\gamma f(\alpha(x)),
    \tilde{F}_\gamma p_2(\alpha(y))) + d_\ominus(\tilde{F}_\gamma
    p_2(\alpha(y)), \tilde{F}_\gamma f(\alpha(y))). \]
  
  The first summand is smaller or equal than $\epsilon$ and it remains
  to show that the second summand equals $0$. We show by contradiction
  that $p_2\le f$ and thus we obtain $0$ via the non-expansiveness of
  the predicate lifting wrt.\ $d_\ominus^\infty$
  (cf. Definition~\ref{def:properties-ev-fct}). So assume that
  $p_2\not\le f$ and we argue that D can not win. There is a state
  $x'$ such that $p_2(x') > f(x')$. S can then at Step~3 choose $p_2$
  and $x'$. Now D chooses at Step~4 $y'$ with $ p_2(x')\leq f(y') $
  and $\epsilon' = f(y') - p_2(x') $. In that case D has no winning
  strategy: $f(y') - f(x')\le d_\alpha^G(x',y')$ since $f$ is
  non-expansive and $f(x') - p_2(x') < 0$, which implies
  $\epsilon' = f(y')-f(x') + f(x') - p_2(x') < d_\alpha^G(x',y') + 0 =
  d_\alpha^G(x',y')$.  And this is a contradiction, since D does not
  have a winning strategy for $(x',y',\epsilon')$.
\end{proof}

\subsubsection{Spoiler Strategy for the Metric Case}

\ThmWinningStrategySpoilerMetric*

\begin{proof}
  Each step described in the strategy yields a smaller formula by
  structural induction. Hence the game will eventually terminate.
  
  We only have to consider the case $\phi = [\gamma]\psi$ in more detail
  and to show (by contradiction) that S can make a valid move in
  Step~3 by proving that the predicate $p_2$ chosen by D in Step~2
  must satisfy $p_2\not\le \llbracket\psi\rrbracket$.
  
  Hence assume that $p_2\le \llbracket\psi\rrbracket$. First observe
  that 
  \[ d_\ominus(\tilde{F}_\gamma p_2(\alpha(y)), \tilde{F}_\gamma
    \llbracket\psi\rrbracket(\alpha(y))) = 0 \] using the
  non-expansiveness of the predicate lifting wrt.\ $d_\ominus^\infty$
  (cf. Definition~\ref{def:properties-ev-fct}).  Using the
  triangle inequality, we have
  \begin{eqnarray*}
    & & d_\ominus(\tilde{F}_\gamma
    \llbracket\psi\rrbracket(\alpha(x)),
    \tilde{F}_\gamma p_2(\alpha(y))) \\
    & \ge & d_\ominus(\tilde{F}_\gamma
    \llbracket\psi\rrbracket(\alpha(x)), \tilde{F}_\gamma
    \llbracket\psi\rrbracket(\alpha(y))) -
    \underbrace{d_\ominus(\tilde{F}_\gamma p_2(\alpha(y)),
      \tilde{F}_\gamma 
      \llbracket\psi\rrbracket(\alpha(y)))}_{=0} \\
    & = &
    d_\ominus(\llbracket\phi\rrbracket(x),\llbracket\phi\rrbracket(y))
    \\
    & > & \epsilon
  \end{eqnarray*}
  which means that D can not play $p_2$ in Step~2 and we have a
  contradiction.
\end{proof}


\end{document}